\documentclass[twocolumn,prd,aps,nofootinbib,showpacs]{revtex4}
\usepackage{multirow}
\usepackage{graphicx}
\usepackage[dvips]{color}
\usepackage{amssymb,amsmath}

\begin{document}

\title{Lattice study on $\pi K $ scattering with moving wall source}

\author{Ziwen Fu }
\affiliation{
Key Laboratory of Radiation Physics and Technology {\rm (Sichuan University)},
Ministry of Education; \\
Institute of Nuclear Science and Technology, Sichuan University,
Chengdu 610064, P. R. China.
}


\begin{abstract}
The $s$-wave pion-kaon ($\pi K$) scattering lengths at zero momentum are calculated
in lattice QCD with sufficiently light $u/d$ quarks and
strange quark at its physical value by the finite size formula.
The light quark masses correspond to $m_\pi = 0.330 - 0.466$ GeV.
In the ``Asqtad'' improved staggered fermion formulation,
we measure the $\pi K$ four-point correlators
for both isospin $I=1/2$ and $3/2$ channels,
and analyze the lattice simulation data at the next-to-leading order
in the continuum three-flavor chiral perturbation theory,
which enables us a simultaneous extrapolation of
$\pi K$ scattering lengths at physical point.
We adopt a technique with the moving wall sources without gauge fixing
to obtain the substantiable accuracy, moreover,
for $I = 1/2$ channel, we employ the variational method
to isolate the contamination from the excited states.
Extrapolating to the physical point yields the scattering lengths as
$m_\pi a_{3/2} = -0.0505(19)$ and  $m_\pi  a_{1/2} = 0.1827(37)$
for $I=3/2$ and $1/2$ channels, respectively.
Our simulation results for $\pi K$ scattering lengths
are in agreement with the experimental reports and theoretical predictions,
and can be comparable with other lattice simulations.
These simulations are carried out with MILC $N_f = 2+1$ flavor
gauge configurations at lattice spacing $a \approx 0.15$~fm.
\end{abstract}

\pacs{12.38.Gc}
\maketitle

\section{Introduction}
Pion-kaon ($\pi K$) scattering at low energies is
the simplest reactions including a strange quark,
and it allows for an explicit exploration of
the three-flavor structure of the low-energy hadronic interactions,
which is not directly probed in the $\pi\pi$ scattering.
The measurement of $\pi K$ scattering lengths is one of the cleanest processes
and a decisive test for our understanding of the chiral $\rm SU(3)$
symmetry breaking of the quantum chromodynamics (QCD).
In the present study, we will concentrate on
the $s$-wave scattering lengths of $\pi K$ system,
which have two isospin eigenchannels ($I=3/2, 1/2$) in the isospin limit,
and the low-energy interaction is repulsive for $I = 3/2$ channel,
and attractive for $I = 1/2$ case, respectively.

Experimentally, $\pi K$ scattering lengths are obtained through
$\pi K$ scattering phases using the Roy-Stainer equations.
The experiments at low energies are an important method
in the study of the interactions among
mesons~\cite{matison74:Kpi_exp_a,johannesson73:Kpi_exp_b,shaw80:Kpi_exp_c},
and these experiments have reported that
the $s$-wave scattering length ($a_0$) in the $I=3/2$ channel,
$m_{\pi} a_{3/2}$ has a small negative value, namely, $-0.13\sim -0.05$.
Moreover, the on-going experiments  proposed by
the DIRAC collaboration~\cite{dirac} to examine $\pi K$ atoms
will provide the direct measurements or
constraints on $\pi K$ scattering lengths.

At present, theory predicts $\pi K$ scattering lengths
with a precision of about $10\%$,
and it will be significantly improved in the near future.
Through the scalar form factors in semi-leptonic pseudo-scalar-to-pseudo-scalar
decays, Flynn et al.~\cite{Fly07} extracted the $\pi K$ scattering length
in the $I=1/2$ channel as $m_{\pi} a_{1/2}$ = +0.179(17)(14).
Three-flavor Chiral Perturbation Theory ($\chi$PT)~\cite{Bernard:1990kw,Ber91b,Kubis:2001bx,Bue03}
has been used to predict the scattering lengths
in the study of the low-energy $\pi K$ scattering,
and small negative value was claimed as $m_{\pi} a_{3/2} = -0.129 \sim -0.05$.
However, if the scattering hadrons contain strange quarks,
$\chi$PT predictions usually suffer from considerable
corrections due to the chiral $\rm SU(3)$ flavor symmetry breaking,
as compared with the case of the $\pi\pi$ scattering.
Therefore, a lattice QCD calculation is needed to
offer an alternative important consistent check
of the validity of $\chi$PT in the presence of the strange quarks.

To date, four lattice studies of $\pi K$ scattering length have been
reported~\cite{Miao:2004gy, Beane:2006gj, Nagata:2008wk, Sasaki:2010zz}.
The first lattice calculation of $\pi K$ scattering length in $I=3/2$ channel
was explored by Miao et al.~\cite{Miao:2004gy} using the quenched approximation,
and the value of $m_\pi a_{3/2}$ was found to be $-0.048$.
The first fully-dynamical calculation using $N_f = 2+1$ flavors of
the Asqtad-improved~\cite{Orginos:1998ue,Orginos:1999cr}
staggered sea quark~\cite{Bernard:2010fr,Bazavov:2009bb}
was carried out~\cite{Beane:2006gj} to calculate
the $I=3/2$ scattering length for $m_\pi = 0.29-0.60$ GeV,
and further indirectly evaluate the $I=1/2$ scattering length
on the basis of $\chi$PT.
They obtained a small negative value of $m_{\pi} a_{3/2} = -0.0574$
for $I=3/2$ channel and a positive value of
$m_{\pi} a_{1/2} = 0.1725$ for $I=1/2$ channel, respectively.
Nagata et al. fulfilled first direct lattice calculation
on $I=1/2$ channel~\cite{Nagata:2008wk} using the quenched approximation.
They investigated all quark diagrams contributing to both isospin eigenstates,
and found that the scattering amplitudes can be expressed as
the combinations of only three diagrams in the isospin limit.
This work greatly inspires us to study $\pi K$ scattering.
However, they did not observe the repulsive interaction
even for $I=3/2$ channel at their simulation points,
and their lattice calculations are relatively cheaper.
Sasaki et al. observed the correct repulsive interaction
for $I=3/2$ channel and attractive for $I=1/2$ case,
and they obtained the scattering lengths of $m_{\pi} a_{3/2} = -0.0500(68)$
and $m_{\pi} a_{1/2} = 0.154(28)$  for the $I=3/2$ and
$1/2$ channels, respectively~\cite{Sasaki:2010zz}.
Moreover, to isolate the contamination from the excited states,
they construct a $2 \times 2$ matrix of
the time correlation function and diagonalize it~\cite{Sasaki:2010zz},
this method will guide us to study $\pi K$ scattering
for $I=1/2$ channel in a correct manner.
It should be stressed that, to reduce the computational cost,
they employed a technique with a fixed kaon sink operator
for the calculation of $\pi K$ scattering length
for $I = 1/2$ channel and then an exponential factor
is introduced to drop the unnecessary $t$-dependence appearing
due to the fixed kaon sink time~\cite{Sasaki:2010zz}.
In this work, we will improve this technique
by using a ``moving'' wall source without gauge fixing
where the exponential factor is not needed any more.
Thus, there is no satisfactory direct lattice calculation
for $I=1/2$ channel until now.

In the present study, we will use the MILC gauge configurations
generated in the presence of $N_f = 2+1$ flavors of the Asqtad
improved~\cite{Orginos:1998ue,Orginos:1999cr} staggered dynamical sea quarks~\cite{Bernard:2010fr,Bazavov:2009bb}
to study the $s$-wave $\pi K$ scattering lengths
for both $I=1/2$ and $3/2$ channels.
Inspired by the exploratory study of $\pi \pi$ scattering
for $I=0$ channel in Ref.~\cite{Kuramashi:1993ka},
we will adopt almost same technique but with moving kaon wall source operator
without gauge fixing for $I = 1/2$ and $3/2$ channels to
obtain the reliable accuracy.
We calculated all the three diagrams categorized in Ref.~\cite{Nagata:2008wk},
and observed a clear signal of attraction for $I=1/2$ channel and
that of repulsion for $I=3/2$ case.
Moreover, for $I = 1/2$ channel, we employ the variational method
to isolate the contamination from the excited states.
Most of all, we only used the lattice simulation data of
our measured $\pi K$ scattering lengths
for both isospin eigenstates to simultaneously extrapolate
toward the physical point using the continuum three-flavor $\chi$PT
at the next-to-leading order.
Our lattice simulation results of the scattering lengths
for both isospin eigenchannels are in accordance with
the experimental reports and theoretical predictions,
and can be comparable with other lattice simulations.

This article is organized as follows.
In Sec.~\ref{sec:Methods} we describe
the formalism for the calculation of $\pi K$ scattering lengths including the
L\"uscher's formula~\cite{Luscher:1991p2480,Luscher:1990ck,Lellouch:2001p4241}
and our computational technique of the modified wall sources
for the measurement of $\pi K$ four-point functions.
In Sec.~\ref{sec:latticeCal} we will show the simulation parameters
and our concrete lattice calculations.
We will present our lattice simulation results in Sec.~\ref{sec:Results},
and arrive at our conclusions and outlooks in Sec.~\ref{sec:conclude}.

\section{Method of measurement}
\label{sec:Methods}
In this section, we will briefly review the formulas of the $s$-wave
scattering length from two-particle energy in a finite box,
with emphasis on the formulae for isospin $I=1/2$ $\pi K$ system.
Also we will present the detailed procedure
for extracting the the energies of $\pi K$ system.
Here we follow the original derivations and notations
in Refs.~\cite{Nagata:2008wk,Sharpe:1992pp,Kuramashi:1993ka,
Fukugita:1994na,Fukugita:1994ve}.

\subsection{ $\pi K$ four-point functions}
\label{SubSec:pK4pFunc}
Let us consider the $\pi K$ scattering of one Nambu-Goldstone pion and
one Nambu-Goldstone kaon in the Asqtad-improved staggered dynamical fermion formalism.
Using operators $O_\pi(x_1), O_\pi(x_3)$ for pions at points $x_1, x_3$,
and operators $O_K(x_2), O_K(x_4)$ for kaons at points $x_2, x_4$, respectively,
with the pion and kaon interpolating field operators defined by
\begin{eqnarray}
{\cal O}_{\pi^+}({\mathbf{x}},t) &=&
- \overline{d}({\mathbf{x}},t)\gamma_5 u({\mathbf{x}},t) \,, \cr
{\cal O}_{\pi^0}({\mathbf{x}},t) &=&
\frac{1}{\sqrt{2}}
[\overline{u}({\mathbf{x}},t)\gamma_5 u({\mathbf{x}},t) -
 \overline{d}({\mathbf{x}},t)\gamma_5 d({\mathbf{x}},t) ] \,, \cr
{\cal O}_{K^0}({\mathbf{x}},t)   &=&
\overline{s}({\mathbf{x}},t)\gamma_5 d({\mathbf{x}}, t) \,, \cr
{\cal O}_{K^+}({\mathbf{x}},t)   &=&
 \overline{s}({\mathbf{x}},t)\gamma_5 u({\mathbf{x}},t) \,,
\end{eqnarray}
we then represent the $\pi K$ four-point functions as
\begin{equation}
C_{\pi K}(x_4,x_3,x_2,x_1) =
\bigl< {\cal O}_K(x_4) {\cal O}_{\pi}(x_3) {\cal O}_K^{\dag}(x_2) {\cal O}_{\pi}^{\dag}(x_1)\bigr> \,,
\end{equation}
where $\langle \cdots\rangle$ represents the expectation value of the path integral,
which we evaluate using the lattice QCD simulations.

After summing over spatial coordinates
${\bf{x}}_1$, ${\bf{x}}_2$, ${\bf{x}}_3$ and ${\bf{x}}_4$,
we obtain the $\pi K$ four-point function in the zero-momentum state,
\begin{equation}
\label{EQ:4point_pK}
C_{\pi K}(t_4,t_3,t_2,t_1) =
\sum_{{\bf{x}}_1} \sum_{{\bf{x}}_2}\sum_{{\bf{x}}_3}\sum_{{\bf{x}}_4}
C_{\pi K}(x_4,x_3,x_2,x_1) \,,
\end{equation}
where $x_1 \equiv ({\bf{x}}_1,t_1)$,  $x_2 \equiv ({\bf{x}}_2,t_2)$,
$x_3 \equiv ({\bf{x}}_3,t_3)$, and $x_4 \equiv ({\bf{x}}_4,t_4)$, and
$t$ stands for the time difference, namely, $t\equiv t_3 - t_1$.

To avoid the complicated Fierz rearrangement of the quark lines,
we choose the creation operators at the time slices
which are different by one lattice time spacing
as is suggested in Ref.~\cite{Fukugita:1994ve}, namely,
we select $t_1 =0, t_2=1, t_3=t$, and $t_4 = t+1$.
In $\pi K$ system, there are two isospin eigenstates,
namely, $I = 3/2$ and $I=1/2$, we construct the $\pi K$ operators
for these isospin eigenchannels as~\cite{Nagata:2008wk}
\begin{eqnarray}
\label{EQ:op_pipi}
{\cal O}_{\pi K}^{I=\frac{1}{2}} (t) &=& \frac{1}{\sqrt{3}}
     \Bigl\{  \sqrt{2}\pi^+(t) K^0(t+1) - \pi^{0}(t) K^{+}(t+1)  \Bigl\} \,, \cr
{\cal O}_{\pi K}^{I=\frac{3}{2}} (t) &=& \pi^{+}(t) K^{+}(t+1) \,,
\end{eqnarray}
where
\begin{eqnarray}
{\cal O}_{K^0}(t)   &=& \sum_{\bf{x}} {\cal O}_{K^0}({\mathbf{x}},t),   \quad
{\cal O}_{K^+}(t)    =  \sum_{\bf{x}} {\cal O}_{K^+}({\mathbf{x}},t)    \cr
{\cal O}_{\pi^0}(t) &=& \sum_{\bf{x}} {\cal O}_{\pi^0}({\mathbf{x}},t), \quad
{\cal O}_{\pi^+}(t)  =  \sum_{\bf{x}} {\cal O}_{\pi^+}({\mathbf{x}},t)  .
\end{eqnarray}

\begin{figure}[thb]
\includegraphics[width=8cm,clip]{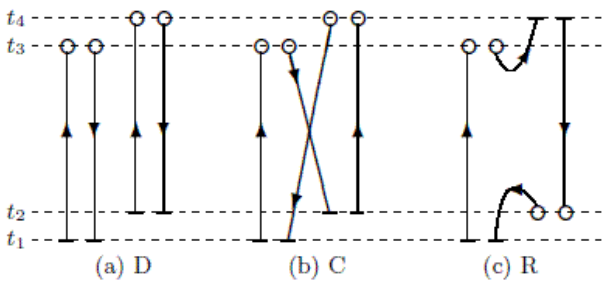}
\vspace{-0.5cm}
\caption{ \label{fig:diagram}
Diagrams contributing to $\pi K$ four-point functions.
Short bars stand for wall sources.
Open circles are sinks for local pion or kaon operators.
The thinker lines represent the strange quark lines.
 }
\end{figure}
If we assume that the $u$ and $d$ quarks have the same mass,
only three diagrams contribute to
$\pi K $ scattering amplitudes~\cite{Nagata:2008wk}.
The quark line diagrams contributing to $\pi K$ four-point function
denoted in Eq.~(\ref{EQ:4point_pK})
are displayed  in Figure~\ref{fig:diagram},
labeling them as direct (D), crossed (C) and rectangular (R), respectively.
The direct and crossed diagrams can be easily evaluated
by constructing the corresponding four-point amplitudes
for arbitrary values of the time slices $t_3$ and $t_4$
using only two wall sources placed at
the fixed time slices $t_1$ and $t_2$.
However, the rectangular diagram (R) requires another quark propagator
connecting the time slices $t_3$ and $t_4$,
which make the reliable evaluation of this diagram extremely difficult.

Sasaki et al. solve this problem through the technique
with a fixed kaon sink operator
to reduce the computational cost~\cite{Sasaki:2010zz}.
Encouraged by the exploratory works of the $\pi \pi $ scattering
at $I=0$ channel in Refs.~\cite{Kuramashi:1993ka,Fukugita:1994ve},
similarly,
we handle this problem by evaluating $T$ quark propagators
on a $L^3 \times T$ lattice, each propagator, which corresponds
to a wall source at the time slice
$ t = 0, \cdots, T-1$, is denoted by
\begin{equation}
\sum_{n''}D_{n',n''}G_t(n'') = \sum_{\mathbf{x}}
\delta_{n',({\mathbf{x}},t)}, \quad 0 \leq t \leq T-1 ,
\end{equation}
where $D$ is the quark matrix for the staggered Kogut-Susskind quark action.
The combination of $G_t(n)$ which we apply for $\pi K$ four-point
functions is schematically shown in Figure~\ref{fig:diagram},
where short bars stand for the position of wall source,
open circles are sinks for local pion or kaon operators,
and the thinker lines represent the strange quark lines.
Likely, the subscript $t$ in the quark propagator $G$
represents the position of the wall source.
$D$, $C$, and $R$, are schematically displayed in Figure~\ref{fig:diagram},
and we can also expressed them in terms of the quark propagators $G$, namely,
\begin{widetext}
\begin{eqnarray}
\label{eq:dcr}
C^D(t_4,t_3,t_2,t_1) &=&
\sum_{ {\bf{x}}_3}\sum_{ {\bf{x}}_4 } \left\langle \mbox{Re} \, \mbox{Tr}
\left[G_{t_1}^{\dag}({\bf{x}}_3, t_3) G_{t_1}({\bf{x}}_3, t_3)
G_{t_2}^{\dag}({\bf{x}}_4, t_4) G_{t_2}({\bf{x}}_4, t_4) \right] \right\rangle,\cr
C^C(t_4,t_3,t_2,t_1) &=&
\sum_{ {\bf{x}}_3}\sum_{{\bf{x}}_4 } \left\langle \mbox{Re} \, \mbox{Tr}
\left[G_{t_1}^{\dag}({\bf{x}}_3, t_3) G_{t_2}({\bf{x}}_3, t_3)
G_{t_2}^{\dag}({\bf{x}}_4, t_4) G_{t_1}({\bf{x}}_4, t_4) \right] \right\rangle,\cr
C^R(t_4,t_3,t_2,t_1) &=&
\sum_{ {\bf{x}}_2}\sum_{ {\bf{x}}_3 } \left\langle \mbox{Re} \, \mbox{Tr}
\left[G_{t_1}^{\dag}({\bf{x}}_2, t_2) G_{t_4}({\bf{x}}_2, t_2)
G_{t_4}^{\dag}({\bf{x}}_3, t_3) G_{t_1}({\bf{x}}_3, t_3) \right] \right\rangle,
\end{eqnarray}
\end{widetext}
where daggers mean the conjugation by the even-odd parity $(-1)^n$
for the staggered Kogut-Susskind quark action, and
$\rm Tr$ stands for the trace over the color index.
The hermiticity properties of the propagator $G$
are used to eliminate the factors of $\gamma^5$.

For $\pi K$ rectangular diagram in Figure~\ref{fig:diagram}(c),
it creates the gauge-variant noise~\cite{Kuramashi:1993ka,Fukugita:1994ve}.
One can reduce the noise by fixing gauge configurations
to some gauge ( e.g., Coulomb gauge), and select
a special wall source to emit only the Nambu-Goldstone pion~\cite{Gupta:1990mr},
however, the gauge non-invariant states may contaminate the $\pi K$ four-point function.
Alternatively, we perform the gauge field average without gauge fixing
since the gauge dependent fluctuations should neatly cancel out
among the lattice configurations. Besides these cancelations,
the summation of the gauge-variant terms over the spatial sites
of the wall source further suppresses the gauge-variant noise.
In our current lattice simulation we found that this method works pretty well.

All three diagrams in Figure~\ref{fig:diagram} are needed to be calculated
to study the $\pi K$ scattering in both $I=1/2$ and $I=3/2$ channels.
Three types of the propagators can be combined to construct the
physical correlation functions for $\pi K$ states with definite isospin.
As it is investigated in Ref.~\cite{Nagata:2008wk},
in the isospin limit,
the $\pi K$ correlation function for $I=3/2$ and $1/2$ channels
can be expressed as the combinations of three diagrams, namely,
\begin{eqnarray}
\label{EQ:phy_I12_32}
C_{\pi K}^{I=\frac{1}{2}}(t) \hspace{-0.15cm} &\equiv& \hspace{-0.15cm}
\left\langle {\cal O}_{\pi K}^{I=\frac{1}{2}} (t) | {\cal O}_{\pi K}^{I=\frac{1}{2}} (0) \right\rangle = D + \frac{1}{2}N_f C - \frac{3}{2}N_f R , \cr
C_{\pi K }^{I=\frac{3}{2}}(t) \hspace{-0.15cm} &\equiv& \hspace{-0.15cm}
\left\langle {\cal O}_{\pi K}^{I=\frac{3}{2}} (t) | {\cal O}_{\pi K}^{I=\frac{3}{2}} (0) \right\rangle =  D - N_f C ,
\end{eqnarray}
where the operator ${\cal O}_{\pi K}^{I}$ denoted in Eq.~(\ref{EQ:op_pipi})
creates a $\pi K$ state with total isospin $I$
and the staggered-flavor factor $N_f$ is inserted to correct for
the flavor degrees of freedom of the Kogut-Susskind staggered fermion~\cite{Sharpe:1992pp}.
For the pion and kaon operators it is most natural to choose
the one in the Nambu-Goldstone channel.
This is the choice for our current study.

To calculate the scattering lengths
for hadron-hadron scattering on the lattice,
or the scattering phase shifts in general,
one usually resorts to L\"uscher's formula which relates the exact energy level
of two hadron states
in a finite box to the scattering phase shift in the continuum.
In the case of $\pi K$ scattering,
the $s$-wave $\pi K$ scattering length in the continuum is defined by
\begin{equation}
\label{eq:exact}
a_0 = \lim_{k\to 0} \frac{\tan\delta_0(k)}{k} .
\end{equation}
$k$ is the magnitude of the center-of-mass scattering momentum
which is related to the total energy $E_{\pi K}^I$ of the $\pi K$ system
with isospin $I$ in a finite box of size $L$  by
\begin{equation}
\label{eq:E_k}
E_{\pi K}^I = \sqrt{m_\pi^2 + k^2} + \sqrt{m_K^2 + k^2} \,,
\end{equation}
where the $m_\pi$ is the pion mass, and $m_K$ is the kaon mass.
We can rewrite Eq.~(\ref{eq:E_k}) to an elegant form as
\begin{equation}
\label{eq:MF_k_e}
k^2  = \frac{1}{4}
\left( E_{\pi K}^I  + \frac{m_\pi^2 - m_K^2}{E_{\pi K}^I } \right)^2 - m_\pi^2 \,.
\end{equation}
In the absence of the interactions between the $\pi$ and $K$ particles,
$k/\tan\delta_0(k) \to \infty $,
and the energy levels occur at momenta $k = 2\pi n/L$, ($n$ is a integer),
corresponding to the single-particle modes in a cubic box.
$\delta_0(k)$ is the $s$-wave scattering phase shift,
which can be evaluated by the L\"uscher's finite size formula~\cite{Luscher:1991p2480,Lellouch:2001p4241},
\begin{equation}
\left( \frac{\tan\delta_0(k)}{k} \right)^{-1} =
\frac{\sqrt{4\pi}}{\pi L}\cdot{\mathcal Z}_{00}\left(1,\frac{k^2}{(2\pi/L)^2}\right) ,
\label{eq:luscher}
\end{equation}
where the zeta function $\mathcal{Z}_{00}(1;q^2)$ is denoted by
\begin{equation}
\label{eq:Z00d}
\mathcal{Z}_{00}(1;q^2) = \frac{1}{\sqrt{4\pi}}
\sum_{{\mathbf n}\in\mathbb{Z}^3} \frac{1}{n^2-q^2}\,,
\end{equation}
here $q=kL/(2\pi)$ is no longer an integer, and $\mathcal{Z}_{00}(1;q^2)$
can be efficiently calculated by the method described in Ref.~\cite{Yamazaki:2004qb}.
We also discussed this technique in Appendix~\ref{appe:zeta},
where we extend this discussion to the case with the negative $q^2$.
In the case of the attractive interaction,
$k^2$ on the bound state has a negative value,
therefore  $k$ is pure imaginary, and $\delta_0(k)$ is
no longer physical scattering phase shift~\cite{Sasaki:2010zz}.
${\mathcal Z}_{00}(1, q^2)$, however, still have a real value even for this case,
hence $\tan\delta_0(k)/k$ obtained by Eq.~(\ref{eq:luscher}) is also real.
If $|k^2|$ is enough small, we can consider $\tan\delta_0(k)/k$ as
the physical scattering length at $\pi k$ threshold~\cite{Sasaki:2010zz}.

The energy $E_{\pi K}^I$ of $\pi K$ system with isospin $I$
can be obtained from $\pi K$ four-point function
denoted in Eq.~(\ref{EQ:phy_I12_32})
with the large $t$. At large  $t$ these correlators
will behave as~\cite{Mihaly:1997,Mihaly:Ph.D}
\begin{eqnarray}
\label{eq:E_pionK}
C_{\pi K}^I(t)  &=&
Z_{\pi K}\cosh\left[E_{\pi K}^I\left(t - \frac{T}{2}\right)\right] +\cr
&&\hspace{-0.85cm}
(-1)^t Z_{\pi K}^{\prime}\cosh\left[E_{\pi K}^{I \prime} \left(t-\frac{T}{2}\right)\right] + \cdots.
\end{eqnarray}
where $E_{\pi K}^I$ is the energy of the lightest $\pi K$ state with isospin $I$.
The terms alternating in sign are a peculiarity of the
Kogut-Susskind formulation of the lattice fermions and correspond to
the contributions from intermediate states with
opposite parity~\cite{Mihaly:1997,Mihaly:Ph.D}.
The ellipsis suggests the contributions from excited states
which are suppressed exponentially.

We should bear in mind that,
for the staggered Kogut-Susskind quark action,
there are further complications in itself
stemming from the non-degeneracy of pions and kaons
in the Goldstone and other channels at a finite lattice spacing.
Briefly speaking, the contributions of non-Nambu-Goldstone
pions and kaons in the intermediate states is exponentially
suppressed for large times due to their heavier masses
compared to these of the Nambu-Goldstone pion
and kaon~\cite{Sharpe:1992pp,Kuramashi:1993ka,Fukugita:1994ve}.
Thus, we suppose that $\pi K$ interpolator does not couple significantly
to other $\pi K$  tastes, and neglect this systematic errors.

In our concrete calculation, we calculated the pion mass $m_\pi$
and kaon mass $m_K$ through the methods discussed by
the MILC collaboration in Refs.~\cite{Bernard:2001av,Aubin:2004wf}
in our previous study~\cite{fzw:2011cpc12}.
In this work we evaluate total energy $E_{\pi K}^I$ of $\pi K$ system
with isospin $I$  from Eq.~(\ref{eq:E_pionK}).

In the current study we also evaluate the energy shift
$\delta E_I = E_{\pi K}^I -  (m_\pi + m_K)$
from the ratios
\begin{equation}
\label{EQ:ratio}
R^X(t) = \frac{ C_{\pi K}^X(0,1,t,t+1) }
{ C_\pi (0,t) C_K(1,t+1) },
\quad  X = D, C, \ {\rm and} \ R \,,
\end{equation}
where $C_\pi (0,t)$ and $C_K (1,t+1)$ are
the pion and kaon two-point functions, respectively.
Considering Eq.~(\ref{EQ:phy_I12_32}),
we can write the amplitudes which project out the $I=1/2$ and $3/2$
isospin eigenstates as
\begin{eqnarray}
\label{EQ:proj_I0I2}
R_{I=\frac{1}{2}}(t) &=&
R^D(t) + \frac{1}{2} N_f R^C(t) - \frac{3}{2} N_f R^R(t) \,, \cr
R_{I=\frac{3}{2}}(t) &=& R^D(t) - N_f R^C(t) \,.
\end{eqnarray}

Following the discussions in Ref.~\cite{Sharpe:1992pp},
we now then can extract the energy shift $\delta E_I$ from the ratios
\begin{equation}
\label{eq:extraction_dE}
R_I(t) = Z_I e^{-\delta E_I t} + \cdots ,
\end{equation}
where $Z_I$ stands for wave function factor,
which is the ratio of two amplitudes from the $\pi K$ four-point function and
the square of the pion two-point correlator and the kaon two-point correlator,
and the ellipsis indicates the terms suppressed exponentially.
In $R_I(t)$, some of the fluctuations which
contribute to both the two-point and four-point
correlation functions neatly cancel out,
hence, improving the quality of the extraction of the energy shift
as compared with what we can obtain from
an analysis through the individual correlation functions~\cite{Beane:2006gj}.

For $I=3/2$ channel, we can use Eq.~(\ref{eq:E_pionK})
or Eq.~(\ref{eq:extraction_dE}) to extract the energy shifts $\delta E$.
We have numerically compared the fitting values from two methods,
and found well agreement within statistical errors.
In fact, using Eq.~(\ref{eq:extraction_dE}) to
extract the energy shift $\delta E$ has been extensively employed
for the study of $\pi K$ scattering at $I=3/2$ case in Ref.~\cite{Beane:2006gj}.
Hence, in this work, we will only present the energy shifts $\delta E$
calculated by Eq.~(\ref{eq:extraction_dE}), and then its
corresponding scattering lengths.

On the other hand, for $I=1/2$ channel,
the presence of the kappa resonance is clearly shown in the low energy~\cite{Sasaki:2010zz},
and therefore it should be necessary for us to separate the ground state contribution
from the contamination stemming from the excited states to
achieve  the reliable scattering length
as it is investigated in Ref.~\cite{Sasaki:2010zz}.
For this purpose, we will construct a $2 \times 2$ correlation matrix
of the time correlation function and diagonalize it
to extract the energy of the ground state.

\subsection{Correlation matrix}
\label{SubSec:Correlation_matrix }
For $I=1/2$ channel, to separate the contamination from the excited states,
we construct a matrix of the time correlation function,
\begin{equation}
C(t) = \left(
\begin{array}{ll}
\langle 0 | {\cal O}_{\pi K}^\dag(t)  {\cal O}_{\pi K}(0) | 0 \rangle &
\langle 0 | {\cal O}_{\pi K}^\dag(t)  {\cal O}_\kappa(0)  | 0 \rangle
\vspace{0.3cm}\\
\langle 0 | {\cal O}_\kappa^\dag(t)   {\cal O}_{\pi K}(0) | 0 \rangle &
\langle 0 | {\cal O}_\kappa^\dag(t)   {\cal O}_\kappa(0)  | 0 \rangle
\end{array} \right) ,
\label{eq:CorrMat}
\end{equation}
where ${\cal O}_\kappa(t)$ is an interpolating operator for the $\kappa$ meson
with zero momentum, and ${\cal O}_{\pi K}(t)$ is
an interpolating operator for $\pi K$ system
which is extensively discussed in section~\ref{SubSec:pK4pFunc}.
The interpolating operator ${\cal O}_\kappa$
employed here is exactly the same as these in our previous studies
in Refs.~{\cite{fzw:2011cpc12,Fu:2011xb,Fu:2011xw},
the notations adopted here are also the same,
but to make this paper self-contained,
all the necessary definitions will be also presented below.

\subsubsection{$\kappa$ sector}
In our previous studies~\cite{fzw:2011cpc12,Fu:2011xb,Fu:2011xw},
we have presented a detailed procedure to measure kappa correlator
$\langle 0 | \kappa^\dag(t) \kappa (0) | 0 \rangle$.
To simulate the correct number of quark species,
we use the fourth-root trick~\cite{Aubin:2003mg}, which automatically
performs the transition from four tastes to one taste per flavor
for staggered fermion at all orders.
We employ an interpolation operator with isospin $I=1/2$
and $J^{P}=0^{+}$ at the source and sink,
\begin{eqnarray}
{\cal O}(x)  \equiv
\frac{1}{\sqrt{n_r}} \sum_{a, g}\bar s^a_g( x ) u^a_g(x)\,,
\end{eqnarray}
where $g$   is the indices of the taste replica,
      $n_r$ is the number of the taste replicas,
      $a$   is the color indices,
      and we omit the Dirac-Spinor index.
The time slice correlator for the $\kappa$ meson
in the zero momentum state can be evaluated by
\begin{equation}
C_\kappa(t) =
\frac{1}{n_r}
\sum_{ {\mathbf{x} }, a, b} \sum_{g, g^\prime} \,
\left\langle \bar s^{b}_{g^\prime}({\mathbf{x}}, t)
u^{b}_{g^\prime}( {\mathbf{x} }, t) \
\bar u^{a}_{g }({\mathbf 0}, 0) s^{a}_{g }({\mathbf 0}, 0)
\right\rangle , \nonumber
\label{EQ.kappa}
\end{equation}
where ${\mathbf 0}, \mathbf{x}$ are
the spatial points of the $\kappa$ state at source and sink, respectively.
After performing Wick contractions of fermion fields,
and summing over the taste index,
for the light $u$ quark Dirac operator $M_u$
and the $s$ quark Dirac operator $M_s$,
we obtain~\cite{fzw:2011cpc12}
\begin{equation}
C_\kappa(t)) = \sum_{\bf x}(-)^x \left\langle
\mbox{Tr}[M^{-1}_u({\bf x},t;0,0)M^{-1^\dag}_s( {\bf x},t;0,0)]
\right\rangle \,,
\label{EQ_kappa}
\end{equation}
where $\mbox{Tr}$ is the trace over the color index,
and $x=({\bf{x}},t)$ is the lattice position.

For the staggered quarks, the meson propagators
have the generic single-particle form,
\begin{equation}
\label{sfits:ch7}
{\cal C}(t) =
\sum_i A_i e^{-m_i t} +
\sum_i A_i^{\prime}(-1)^t e^{-m_i^\prime t}  +(t \rightarrow N_t-t)\,,
\end{equation}
where the oscillating terms correspond to a particle with opposite parity.
For $\kappa$ meson correlator,
we consider only one mass with each parity in Eq.~(\ref{sfits:ch7}),
namely, in our concrete calculation,
our operator is the state with spin-taste assignment $I \otimes I$
and its oscillating term with $\gamma_0\gamma_5 \otimes \gamma_0\gamma_5$~\cite{fzw:2011cpc12}.
Thus, the $\kappa$ correlator was fit to the following physical model,
\begin{equation}
\label{eq:kfit}
C_{\kappa}(t) = b_{\kappa}e^{-m_{\kappa}t} +
b_{K_A}(-)^t e^{-M_{K_A}t} + (t \rightarrow N_t-t)\,,
\end{equation}
where $b_{K_A}$ and $b_{\kappa}$ are two overlap factors.
In Figure~\ref{fig:G_t}, we clearly note this oscillating term.

We should bear in mind that, for the staggered Kogut-Susskind quark action,
our $\kappa$ interpolating operator couples to various tastes
as we examined the scalar $a_0$ and $\sigma$ mesons
in Refs.~\cite{Bernard:2007qf,Bernard:2006gj}, where we investigated
two-pseudoscalar intermediates states (namely, bubble contribution).
In Ref.~\cite{Fu:2011xb},
we investigated the extracted $\kappa$ masses
with and without bubble contribution for kappa correlators.
We found that there are only about $2\% \thicksim 5\%$ differences in masses,
although the bubble contributions are dominant
in the $\kappa$ correlators at large time region.
Thus, in the current study,
we can reasonably assume that the $\kappa$ interpolator does not couple remarkably
to other tastes, and ignore this systematic errors
for the $\kappa$ sector~\cite{Fu:2011xw}.

\subsubsection{ Off-diagonal sector}
The calculations of the off-diagonal elements in correlation matrix $C(t)$
in Eq.~(\ref{eq:CorrMat}), namely,
$\langle 0 | {\cal O}_{\pi K}^\dag(t)  {\cal O}_\kappa(0)  | 0 \rangle$ and
$\langle 0 | {\cal O}_\kappa^\dag(t)   {\cal O}_{\pi K}(0)$
are exactly the same as these in our previous study for non-zero momenta
in Ref.~{\cite{Fu:2011xw},
the notations adopted here are also the same,
but to make this paper self-contained,
all the necessary definitions will be also presented below.

To avoid the complicated Fierz rearrangement of the quark lines,
we choose the creation operators at the time slices
which are different by one lattice time spacing
as suggested in Ref.~\cite{Fukugita:1994ve}, namely,
we select $t_1 =0, t_2=1$, and $t_3=t$
for $\pi K \to \kappa$ three-point function,
and choose $t_1 =0, t_2=t$, and $t_3=t+1$
for $\kappa \to \pi K$ three-point function.

\begin{figure}[th]
\begin{center}
\includegraphics[width=6cm,clip]{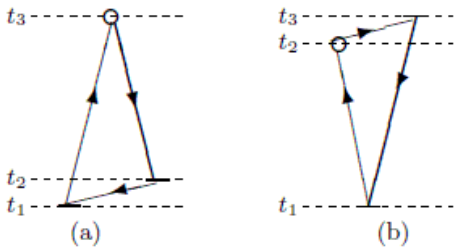}
\end{center}
\vspace{-0.5cm}
\caption{ \label{fig:3diagram}
Diagrams contributing to $\pi K \to \kappa$ and
$\kappa \to \pi K$ three-point functions.
Short bars stand for wall sources.
The thinker lines represent the strange quark lines.
(a) Quark contractions of $\pi K \to \kappa$,
where open circle is sink for local kappa operator.
(b) Quark contractions of $\kappa \to \pi K$,
where open circle is sink for local pion operator.
}
\end{figure}

The quark line diagrams contributing to the $\kappa \to \pi K$ and
$\pi K \to \kappa$  three-point function are plotted
in Figure~\ref{fig:3diagram}(a) and Figure~\ref{fig:3diagram}(b), respectively,
where short bars stand for the position of wall source,
open circles are sinks for local pion or kaon operators,
and the thinker lines represent the strange quark lines.
Likely, the subscript $t$ in the quark propagator $G$
represents the position of the wall source.

The $\pi K \to \kappa$  three-point function can be easily evaluated
by constructing the corresponding three-point amplitudes
for arbitrary values of the time slice $t_3$
using only two wall sources placed at
the fixed time slices $t_1$ and $t_2$.
However, the calculation of $\kappa \to \pi K$ three-point function
is almost the same difficult as that of
the rectangular diagram for $\pi K$ four-point correlator function,
since it requires another quark propagator
connecting time slices $t_2$ and $t_3$.
The $\kappa \to \pi K$ and $\pi K \to \kappa$  three-point functions
are schematically shown in Figure~\ref{fig:3diagram},
and we can also express them in terms of the quark propagators $G$, namely,
\begin{widetext}
\begin{eqnarray}
\label{eq:dcr3}
C_{\pi K \to \kappa} (t_3,t_2,t_1) &=&
\sum_{ {\mathbf{x}}_3, {\mathbf{x}}_1} \left\langle \mbox{Re} \, \mbox{Tr}
[G_{t_1}({\mathbf{x}}_3, t_3) G_{t_2}^{\dag}({\mathbf{x}}_3, t_3)
 G_{t_2}^{\dag}({\mathbf{x}}_1, t_1) ] \right\rangle,\cr
C_{\kappa \to \pi K}(t_3,t_2,t_1) &=&
\sum_{ {\mathbf{x}}_2, {\mathbf{x}}_1} \left\langle \mbox{Re} \, \mbox{Tr}
[G_{t_1}({\mathbf{x}}_2, t_2)
 G_{t_3}^{\dag}({\mathbf{x}}_2, t_2)
 G_{t_3}^{\dag}({\mathbf{x}}_1, t_1) ] \right\rangle.
\end{eqnarray}
\end{widetext}

\subsection{ Extraction of energies }
\label{SubSec:Extraction of energies}
Through calculating the matrix of correlation function $C(t)$
denoted in Eq.~(\ref{eq:CorrMat}),
we can separate the ground state from first excited state in a clean way.
It is very important to map out ``avoided level crossings''
between the $\kappa$ resonance and its decay products (namely, $\pi$ and $K$)
in a finite box volume,
because the first excited state is potentially close to the ground state.
This makes the extraction of the ground state energy unfeasible
if we only utilize a single exponential fit ansatz.
Since we can not predict a priori whether our energy eigenvalues
are near to the resonance region or not,
we find it always safe in practice to adopt the correlation matrix
to analyze our lattice simulation data for isospin $I=1/2$ channel.
To extract the ground state,
we follow the variational method~\cite{Luscher:1990ck}
and construct a ratio of correlation function matrices as
\begin{equation}
M(t,t_R) = C(t) \, C^{-1}(t_R) \,,
\label{eq:M_def}
\end{equation}
with some reference time slice $t_R$~\cite{Luscher:1990ck},
which is assumed to be large enough such that the contributions
to correlation matrix $M(t, t_R)$ from the excited states can be neglected,
and the lowest two eigenstates dominate the correlation function.
The two lowest energy levels can be extracted
by a proper fit to two eigenvalues $\lambda_n (t,t_R)$ ($n=1,2$) of matrix $M(t,t_R)$.
Because here we work on the staggered fermions,
and we can easily prove that $\lambda_n (t,t_R)$ ($n=1,2$)
behaves as~{\cite{Fu:2011xw}
\begin{eqnarray}
\label{Eq:asy}
\lambda_n (t, t_R) &=&  A_n
\cosh\left[-E_n\left(t-\frac{T}{2}\right)\right] + \cr
&&\hspace{-0.85cm} (-1)^t B_n
\cosh\left[-E_n^{\prime}\left(t-\frac{T}{2}\right)\right] \,,
\end{eqnarray}
for a large $t$, which mean $0 \ll t_R < t \ll T/2$ to
suppress the excited states and the unwanted thermal contributions.
This equation explicitly contains an oscillating term.
For the current study, we are only interested in eigenvalue $\lambda_1(t,t_R)$,
here non-degenerate eigenvalues $\lambda_1(t,t_R) > \lambda_2(t,t_R)$ are assumed.
In practice, we found that the oscillating term
in $\lambda_1(t,t_R)$ is not appreciable for some $t_R$,
we can also adopt following simple fitting model~{\cite{Fu:2011xw},
\begin{equation}
\label{Eq:asy_simple}
\lambda_1 (t, t_R) =  A \cosh(-E (t-T/2)) \,,
\end{equation}
and the difference between the fitting models of
Eq.~(\ref{Eq:asy}) and Eq.~(\ref{Eq:asy_simple}) is small.
However, to make our extracted ground energy $E$
for isospin $I=1/2$ channel always reliable,
in this work, we will present the ground energy $E$
calculated by Eq.~(\ref{Eq:asy}),
and then its corresponding $s$-wave scattering lengths.

\section{Lattice calculation}
\label{sec:latticeCal}

\subsection{ Simulation parameters }
\label{SubSec:Simulation Parameters}
We used the MILC lattices with $2+1$ dynamical flavors of the Asqtad-improved
staggered dynamical fermions, the detailed description of the simulation parameters
can be found in Refs.~\cite{Bernard:2010fr,Bazavov:2009bb}.
One thing we must stress that
the MILC configurations are generated using the staggered
formulation of lattice fermions~\cite{Kaplan:1992bt,Shamir:1993zy,Shamir:1998ww}
with rooted staggered sea quarks~\cite{Bernard:2001av}
which are hypercubic-smeared (HYP-smeared)~\cite{Hasenfratz:2001hp,DeGrand:2002vu,DeGrand:2003in,Durr:2004as}.
As it was shown in Refs.~\cite{Renner:2004ck,Edwards:2005kw} that
HYP-smearing gauge links significantly improves the chiral symmetry.

We analyzed $\pi K$ four-point functions on the $0.15$ fm
MILC  lattice ensemble of $450$ $16^3 \times 48$ gauge configurations
with bare quark masses $am_{ud} = 0.0097$
and $am_s = 0.0484$ and bare gauge coupling $10/g^2 = 6.572$,
which has a physical volume approximately $2.5$ fm.
The inverse lattice spacing $a^{-1}=1.358^{+35}_{-13}$ GeV~\cite{Bernard:2010fr,Bazavov:2009bb}.
The mass of the dynamical strange quark is near to its physical value,
and the masses of the $u$ and $d$ quarks are degenerate.
To avoid the contamination from pions and kaons propagating backward in time,
periodic boundary condition is applied to the three spatial directions
while in the temporal direction, Dirichlet boundary condition is imposed,
which reduce the original time extent of $48$ down to $24$,
moreover, it avoids the ``fake effects'' discussed in Ref.~\cite{Sasaki:2010zz}.

\subsection{Sources for isospin $I=1/2$ channel}
To calculate the $\pi K$ correlation functions,
we use the standard conjugate gradient method (CG) to obtain
the required matrix element of the inverse fermion matrix.
The calculation of the correlation function for the rectangular diagrams
naturally requires us to compute the propagators on
all the time slices $ t_s = 0, \cdots, T-1$ of both source and sink,
which requires the calculation of $48$ separate propagators
in our lattice simulations. After averaging the correlator over
all $48$ possible values, the statistics are greatly improved
since we can put the pion source and kaon source at all possible time slices,
namely, the correlator $C_{11}(t)$ is calculated through
\begin{eqnarray}
 C_{11}(t) &=& \left\langle
\left(\pi K\right)(t)\left(\pi K\right)^\dag(0) \right\rangle  \cr
&=&
\frac{1}{T}\sum_{t_s}\left\langle
\left(\pi K\right)(t+t_s)\left(\pi K\right)^\dag(t_s)\right\rangle ,
\end{eqnarray}
The best-effort to generate the propagators on all time slices
enables us to obtain the correlators with high precision,
which is vital to extract the desired scattering phase shifts reliably.

For each time slice, six fermion matrix inversions are
required corresponding to the possible $3$ color choices
for the pion source and kaon source, respectively,
and each inversion takes about one thousand iterations during the CG calculation.
Therefore, all together we carry out $288$ inversions
on a full QCD configuration.
As shown follow, this large number of inversions,
performed on $450$ configurations,
provides the substantial statistics needed to resolve the real parts
of the $I = 1/2$ and $3/2$ amplitudes with reliable accuracy.

In the calculation of the off-diagonal correlator, $C_{21}(t)$,
the quark line contractions results in a three-point diagram.
Since in this three-point diagram the pion field and kaon field
are located at the source time slice $t_s$, $t_s+1$, respectively.
We calculate the off-diagonal correlator $C_{21}(t)$ through
\begin{eqnarray}
C_{21}(t) &=& \left\langle\kappa(t)(\pi K)^\dag(0)\right\rangle \cr
&=&
\frac{1}{T}\sum_{t_s}
\left\langle\kappa(t+t_s)(\pi K)^\dag(t_s)\right\rangle ,
\end{eqnarray}
where, again, we sum the correlator over all time slices $t_s$ and average it.
As for the second off-diagonal correlator $C_{12}(t)$,
the pion field and kaon field are placed
at the sink time slices $t_s+t$  and $t+t_s+1$, respectively,
which make the computation of $C_{12}(t)$ difficult.
However, using the relation $C_{12}(t)=C_{21}^\ast(t)$,
we can obtain the matrix element $C_{12}$ for free.
As it is studied in Ref.~\cite{Aoki:2007rd},
since the sink and source operators are identical
for a large number of configurations, $C(t)$ is a Hermitian matrix.
The $\kappa\to\pi K$ component agrees with $\pi K \to \kappa$ within the error,
but the statistical errors of the matrix element $C_{12}$
should be larger than that of matrix element $C_{21}$ for a large $t$.
Therefore, in the following analyses we substitute matrix element $C_{12}$
by the complex conjugate of matrix element $C_{21}$,
which is not only  to save about $20\%$ computation time,
but also significantly to reduce statistical errors.

For the $\kappa$ correlator, $C_{22}(t)$,
we have measured the point-to-point correlators
with high precision in our previous work~\cite{fzw:2011cpc12}.
Therefore, we can exploit the available propagators
to construct the $\kappa$-correlator
\begin{equation}
C_{22}(t)=\frac{1}{T}\sum_{t_s}
\left\langle\kappa^\dag(t+t_s) \kappa(t_s)\right\rangle ,
\end{equation}
where, also, we sum the correlator over all time slices $t_s$ and average it.

One thing we must stress that,
in the calculation of the correlator $\langle(\pi K)(\pi K)^\dag\rangle$,
we make our best-efforts to reliably measure the rectangular diagram,
since the other two diagrams are relatively easy to evaluate.
We found that the rectangular diagram plays a major role in this correlator.
Therefore, we get it properly for the $\pi K$ sector for isospin $I=1/2$ channel.

In this work, we also measure two-point correlators for pion and kaon, namely,
\begin{eqnarray}
G_\pi(t) &=& \langle 0| \pi^\dag (t) \pi(0) |0\rangle \,, \cr
G_K(t)   &=& \langle 0|   K^\dag (t)   K(0) |0\rangle \,,
\label{eq:Gpi}
\end{eqnarray}
where the $G_\pi (t)$ is correlation function for the pion with zero momentum,
and the $G_K (t)$ is correlation function for the kaon with zero momentum.

\section{Simulation results}
\label{sec:Results}
In our previous work~\cite{fzw:2011cpl},
we have measured the pion and kaon point-to-point correlators.
Using these correlators, we can precisely extract
the pion mass ($m_\pi$) and kaon mass ($m_K$)~\cite{fzw:2011cpl},
which are summarized in Table~\ref{tab:m_pi_K}.
Using the same method discussed in Ref.~\cite{Aubin:2004fs} and the MILC code
for calculating the pion decay constants $f_\pi$,
we precisely extract the pion decay constants $f_\pi$~\cite{Fu:2011bz},
which are in agreement with the previous MILC determinations
at this same lattice ensemble in Ref.~\cite{Bazavov:2009bb}.
Here we used all the $631$ lattice configurations of this ensemble.
We also recapitulated these fitted values in Table~\ref{tab:m_pi_K}.

\begin{table}[h]	
\caption{\label{tab:m_pi_K}
Summary of the pion masses, kaon masses and the pion decay constants.
The third and fourth blocks show pion masses and kaon masses in lattice units,
respectively, and Column five shows the pion decay constants in lattice units.
The second block gives pion masses in GeV,
where the errors are estimated from both the error on the lattice spacing $a$
and the statistical errors in Column three.
}
\begin{ruledtabular}
\begin{tabular}{ccccc}
$am_x$ & $m_\pi({\rm GeV})$ & $a m_\pi$ & $a m_K $ & $a f_{\pi}$  \\
\hline
$0.00970$  & $0.334(6)$  & $0.2459(2)$  & $0.3962(2)$  & $0.12136(29)$ \\
$0.01067$  & $0.350(6)$  & $0.2575(2)$  & $0.3996(2)$  & $0.12264(34)$ \\
$0.01261$  & $0.379(7)$  & $0.2789(2)$  & $0.4066(2)$  & $0.12425(27)$ \\
$0.01358$  & $0.392(7)$  & $0.2890(2)$  & $0.4101(2)$  & $0.12482(32)$ \\
$0.01455$  & $0.406(7)$  & $0.2987(2)$  & $0.4134(2)$  & $0.12600(26)$ \\
$0.01940$  & $0.466(8)$  & $0.3430(2)$  & $0.4300(2)$  & $0.12979(27)$ \\
\end{tabular}
\end{ruledtabular}
\end{table}

\subsection{Diagrams $D, C$, and $R$}
The $\pi K$ four-point functions are calculated
with same lattice configurations using six $u$ valence quarks, namely,
$am_x = 0.0097$, $0.01067$, $0.01261$, $0.01358$, $0.01455$ and $0.0194$,
where $m_x$ is the light valence $u$ quark mass.
They all have the same strange sea quark mass $a m_s=0.0426$,
which is fixed at its physical value~\cite{Bazavov:2009bb}.

In Figure~\ref{fig:ratio} the individual ratios,
which are defined in Eq.~(\ref{EQ:ratio}) corresponding to the diagrams
in Figure~\ref{fig:diagram}, $R^X$ ($X=D, C$ and $R$)
are displayed as the functions of $t$ for $am_x=0.0097$.
We can note that diagram $D$ makes the biggest contribution,
then diagram $C$, and diagram $R$ makes the smallest contribution.
The calculation of the amplitudes for the rectangular diagram
stands for our principal work.
Clear signals observed up to $t = 20$ for the rectangular amplitude
demonstrate that the method of the moving wall source without gauge fixing
used here is practically applicable.
\begin{figure}[h]
\includegraphics[width=8cm,clip]{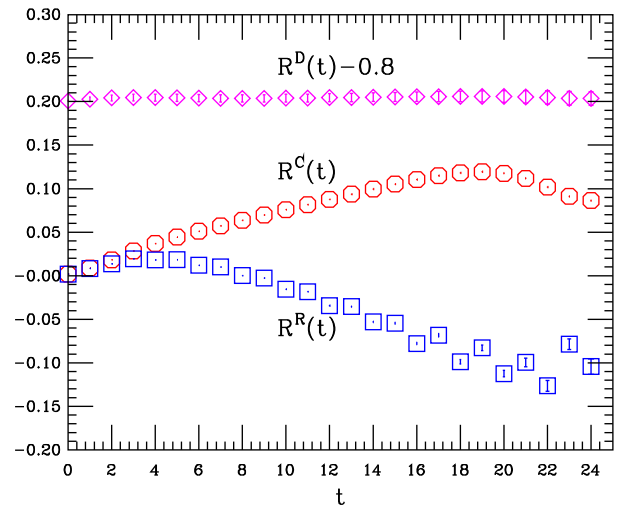}
\caption{(color online).
Individual amplitude ratios $R^X(t)$ of the
$\pi K$ four-point function calculated through the moving wall source
without gauge fixing as the functions of $t$ for $am_x=0.0097$.
Direct diagram (magenta diamonds) shifted by $0.8$,
crossed diagram (red octagons) and rectangular diagram (blue squares).
\label{fig:ratio}
}
\end{figure}

The values of the direct amplitude $R^D$ are quite close to unity,
indicating that the interaction in this channel is very weak.
The crossed amplitude, on the other hand, increases linearly, which
implies a repulsion in $I=3/2$ channel.
After an initial increase up to $t \sim 4$, the rectangular amplitude
exhibits a roughly linear decrease up to $t \sim 20$,
which suggests an attractive force between the pion and kaon in $I=1/2$ channel.
Furthermore, the magnitude of the slope is similar to
that of the crossed amplitude but with opposite sign.
These features are what we eagerly expected from
the theoretical predictions~\cite{Bernard:1990kw,Sharpe:1992pp}.
We can observe that the crossed and rectangular amplitudes have
the same value at $t=0$,
and the close values for small $t$.
Because our analytical expressions in Eq.~(\ref{eq:dcr})
for the two amplitudes coincide at $t=0$,
they should behave similarly until the
asymptotic $\pi K$ state is reached.
\begin{figure}[thb!]
\includegraphics[width=8cm,clip]{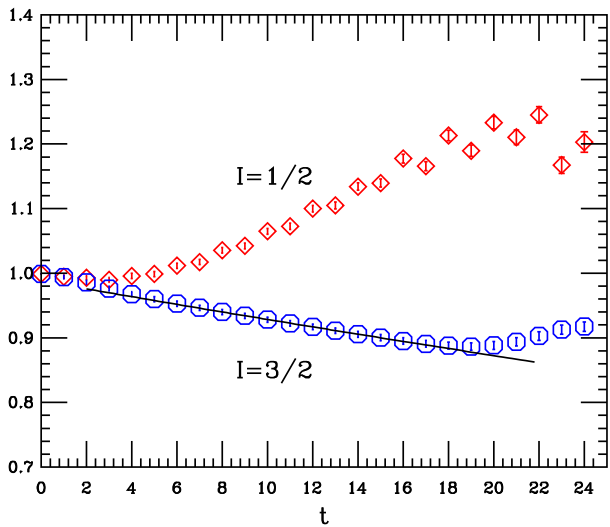}
\caption{(color online).
$R_I(t)$ for $\pi K$ four-point function at zero momenta
calculated by the moving wall source without gauge fixing
as the functions of $t$ for $am_x=0.0097$.
Solid line in $I=3/2$ is exponential fits for $7 \le t \le 16$.
\label{fig:I0I2}
}
\end{figure}

In Figure~\ref{fig:I0I2}, we display the ratio $R_I(t)$ projected
onto the isospin $I = 1/2$ and $3/2$ channels for $am_x=0.0097$,
which are denoted in Eq.~(\ref{EQ:proj_I0I2}).
A decrease of the ratio of $R_{I=3/2}(t)$ indicates  a positive energy shift
and hence a repulsive interaction for $I = 3/2$ channel,
while an increase of $R_{I=1/2}(t)$ suggests a negative energy shift
and hence an attraction for $I = 1/2$ channel.
A dip at $t=3$ for$I = 1/2$ channel
can be clearly observed~\cite{Fukugita:1994ve}.
The systematically oscillating behavior for $I=1/2$ channel
in the large time region is also clearly observed,
which is a typical characteristic of
the Kogut-Susskind formulation of lattice fermions and
corresponds to the contributions from the intermediate states of
the opposite parity~\cite{Mihaly:1997,Mihaly:Ph.D},
this also clearly indicates the existence of the contaminations
from other states rather than the pion-kaon scattering
state~\cite{Sasaki:2010zz}.
Therefore, to isolate the potential contaminations,
we will use the variational method~\cite{Luscher:1990ck}
to analyze the lattice simulation data.
As for $I=3/2$ channel, this oscillating characteristic is not appreciable,
we will use the traditional method, namely,
using Eq.~(\ref{eq:extraction_dE}) to  compute the energy shift
$\delta E$ and then calculate the corresponding scattering length.

\subsection{Fitting analyses for $I=3/2$ channel}
\begin{figure*}[htb]
\begin{center}
\includegraphics[width=8cm]{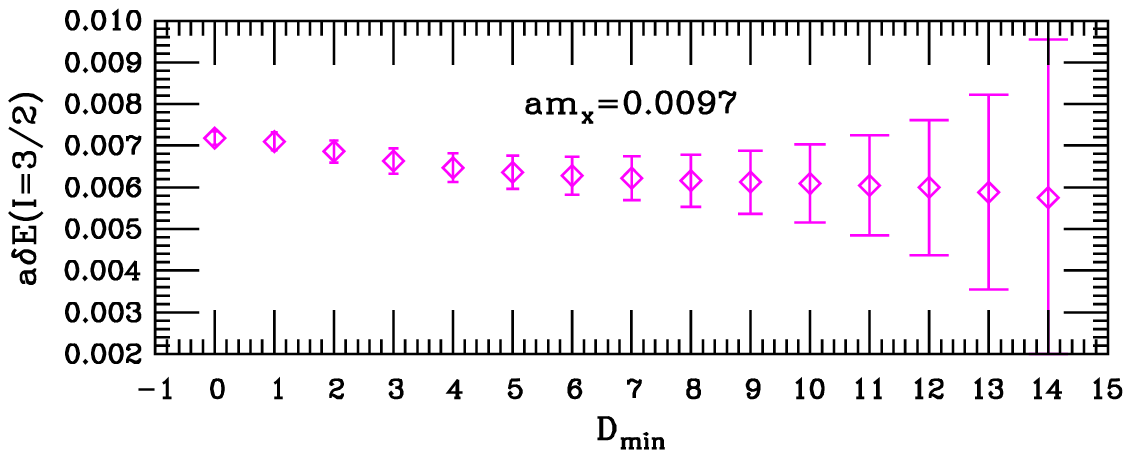}  \hspace{0.6cm}
\includegraphics[width=8cm]{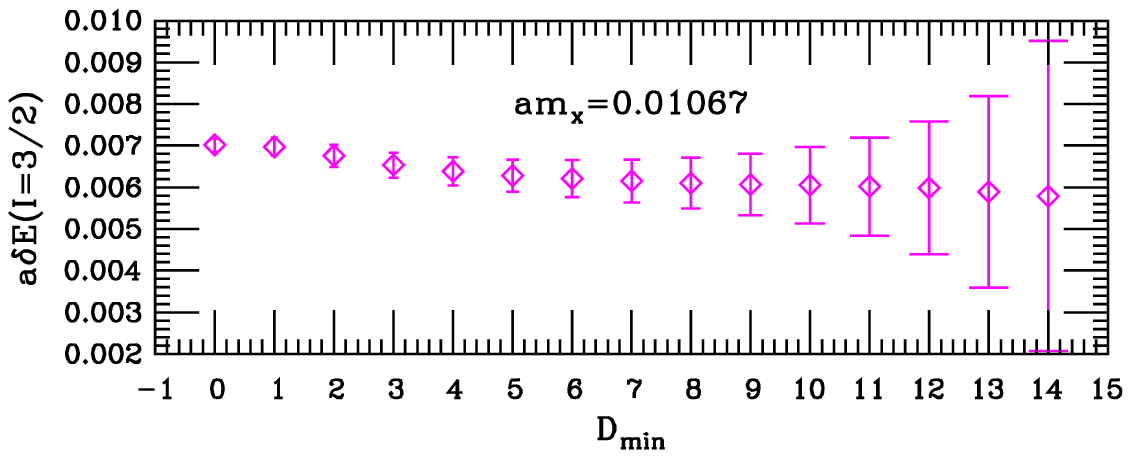}
\includegraphics[width=8cm]{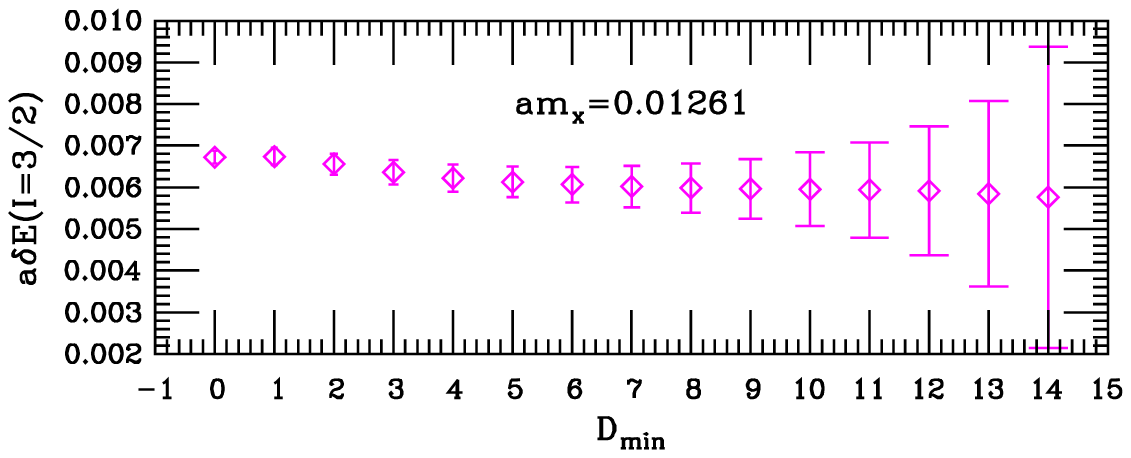}  \hspace{0.6cm}
\includegraphics[width=8cm]{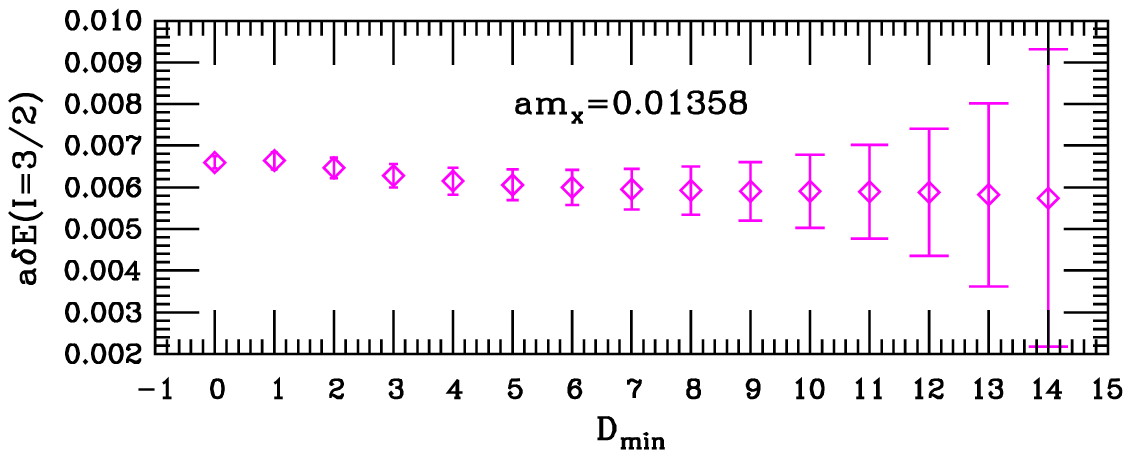}
\includegraphics[width=8cm]{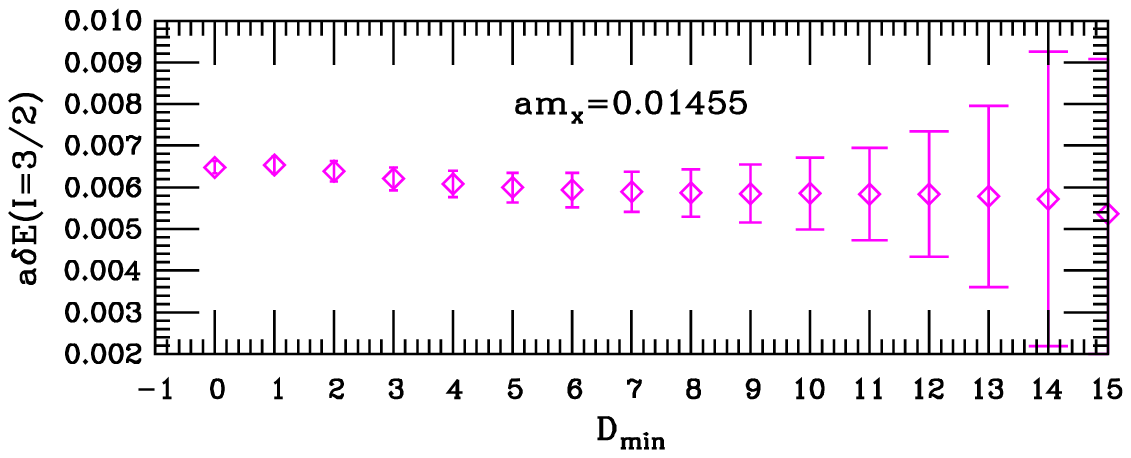}   \hspace{0.6cm}
\includegraphics[width=8cm]{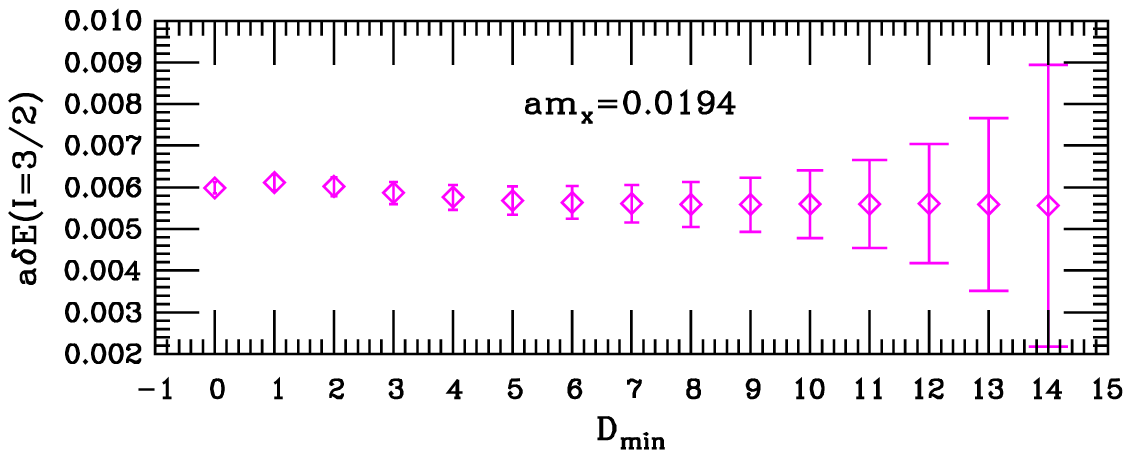}
\end{center}
\vspace{-0.5cm}
\caption{(color online).
The effective $\pi K$ energy shift plots, $a \delta E$
as the functions of the minimum fitting distance $\rm D_{min}$
in the fit for $I=3/2$ channel.
The effective $\pi K$ energy shift plots for $I=3/2$ channel
have only relative small errors within a broad minimum distance
region $5 \le \rm {D_{min}} \le 10$.
\label{fig:eff_eng_shift}
}
\end{figure*}

According to our discussions in Section~\ref{sec:Methods},
in this work, we will make use of Eq.~(\ref{eq:extraction_dE}) to
extract the energy shift $\delta E$ for $I=3/2$ channel.
Then we insert these energy shifts into
the Eqs.~(\ref{eq:exact}) and (\ref{eq:luscher})
to obtain the corresponding $s$-wave scattering lengths.
Therefore, properly extracting the energy shifts is a crucial step
to our final results in this paper.
A convincing way to analyze our lattice simulation data is
with the ``effective energy shift'' plots,
a variant of the effective mass plots,
where the propagators were fit with
varying minimum fitting distances $\rm D_{min}$,
and with the maximum distance $\rm D_{max}$
either at the midpoint of the lattice or
where the fractional statistical errors exceeded about $20\%$
for two successive time slices.
For each valence quark $m_x$,
the effective energy shift plots as a function of
minimum fitting distance  $\rm D_{min}$ for $I=3/2$ channel
are shown in Figure~\ref{fig:eff_eng_shift}.
The central value and uncertainty of each parameter was determined
by the jackknife procedure over the ensemble of gauge configurations.

The energy shifts $a \delta E$ of $\pi K$ system for
$I = 3/2$ channel are extracted from the ``effective energy shift'' plots,
and the energy shifts were selected by looking for a combination of a
``plateau'' in the energy as a function of the minimum distance $\rm D_{min}$,
and a good confidence level (namely, $\chi^2$) for the fit.
We found  that the effective energy shifts for $I=3/2$ channel
have only relative small errors within broad minimum time distance region
$5 \le \rm {D_{min}} \le 10$ and are taken to be quite reliable.

We utilize the exponential physical fitting model in Eq.~(\ref{eq:extraction_dE})
to extract the desired energy shifts for $I=3/2$ channel.
In Figure~\ref{fig:I0I2} we display the ratio $R_I(t)$ projected
onto the $I = 1/2$ and $3/2$ channels for $am_x=0.0097$,
where we can watch the fitted functional form
as compared with the lattice simulation data for $I=3/2$ channel.
For the other five light $u$ valence quarks, we obtain the similar results,
therefore we do not show these ratio $R_I(t)$ plots here.
The fitted values of the energy shifts, $\delta E_I$
in lattice units and wave function factor $Z_I$
for $I=3/2$ channel
are summarized in Table~\ref{tab:energy_shifts}.
The wave function $Z$ factors are pretty close to unity
and the $\chi^2/{\rm dof}$ is quite small for $I=3/2$ channel,
indicating the values of the extracted scattering lengths are substantially reliable.
\begin{table}[h]	
\caption{\label{tab:energy_shifts}
Summary of the lattice simulation results for the energy shifts
in lattice units for $I=3/2$ channel.
The third block shows the energy shifts in the lattice unit,
Column four shows the wave function factors $Z$,
Column five shows the time range for the chosen fit,
and Column six shows the number of degrees of freedom (dof) for the fit.
All errors are calculated from jackknife.
}
\begin{ruledtabular}
\begin{tabular}{cccccl}
Isospin &$a m_x$   & $a\delta E$ & $Z$ & Range  & $\chi^2/{\rm dof}$ \\
\hline
\multirow{6}*{$I=\frac{3}{2}$}
&$0.00970$  & $0.00621(53)$  & $0.9880(59)$  & $7-16$  & $0.0536/8$  \\
&$0.01067$  & $0.00615(52)$  & $0.9893(58)$  & $7-16$  & $0.0395/8$ \\
&$0.01261$  & $0.00602(50)$  & $0.9914(56)$  & $7-16$  & $0.0226/8$ \\
&$0.01358$  & $0.00595(49)$  & $0.9923(55)$  & $7-16$  & $0.0176/8$ \\
&$0.01455$  & $0.00589(48)$  & $0.9930(54)$  & $7-16$  & $0.0142/8$ \\
&$0.01940$  & $0.00561(45)$  & $0.9958(50)$  & $7-16$  & $0.0067/8$ \\
\end{tabular}
\end{ruledtabular}
\end{table}

Now we can insert these energy shifts in Table~\ref{tab:energy_shifts}
into the Eqs.~(\ref{eq:exact}) and (\ref{eq:luscher})
to obtain the corresponding $s$-wave scattering lengths.
The center-of-mass scattering momentum $k^2$ in GeV
calculated by Eq.~(\ref{eq:MF_k_e}),
from which we can easily estimate its statistcal errors.
Once we obtain the values of $k^2$,
the $s$-wave scattering lengths $a_0$ in lattice units
can be obtained through Eqs.~(\ref{eq:exact}) and (\ref{eq:luscher}).
All of these values are summarized in Table~\ref{tab:I_3_2}.
Here we utilize pion masses and kaon masses given in Table~\ref{tab:m_pi_K}.
The errors of the center-of-mass scattering momentum $k$ and
the $s$-wave scattering lengths
come from the statistical errors of
the energy shifts energies $\delta E$,
pion mass $m_\pi$ and kaon mass $m_K$.

\begin{table}[h]	
\caption{\label{tab:I_3_2}
Summary of the lattice simulation results of
the $s$-wave scattering lengths for $I=3/2$ channel.
The third block shows the center-of-mass scattering momentum $k^2$ in GeV,
Column four shows the $s$-wave scattering lengths in lattice units,
and Column five shows the pion mass times scattering lengths.
}
\begin{ruledtabular}
\begin{tabular}{ccccl}
Isospin &$am_x$   &  $k^2$[${\rm GeV}^2$]  & $a_0$  & $m_\pi a_0$\\
\hline
\multirow{6}*{$I=\frac{3}{2}$}
&$0.00970$  & $0.00350(27)$  & $-0.558(55)$  & $-0.137(13) $ \\
&$0.01067$  & $0.00357(28)$  & $-0.569(55)$  & $-0.146(14)$ \\
&$0.01261$  & $0.00366(30)$  & $-0.582(55)$  & $-0.162(15)$ \\
&$0.01358$  & $0.00374(30)$  & $-0.593(56)$  & $-0.171(16)$ \\
&$0.01455$  & $0.00379(34)$  & $-0.600(60)$  & $-0.179(18)$ \\
&$0.01940$  & $0.00396(37)$  & $-0.624(62)$  & $-0.214(21)$ \\
\end{tabular}
\end{ruledtabular}
\end{table}

\subsection{Fitting analyses for $I=1/2$ channel}
In Figure~\ref{fig:G_t}, we show the real parts of the diagonal components
($\pi K\to\pi K$ and $\kappa\to\kappa$)
and the real part of the off-diagonal component $\pi K\to\kappa$
of the correlation function $C(t)$ denoted in Eq.~(\ref{eq:CorrMat}).
Since $C(t)$ is a Hermitian matrix,
we will substitute the off-diagonal component $\kappa\to\pi K$
by $\pi K\to\kappa$ to reduce statistical errors
in the following analyses.

\begin{figure}[h]
\begin{center}
\includegraphics[width=8.0cm]{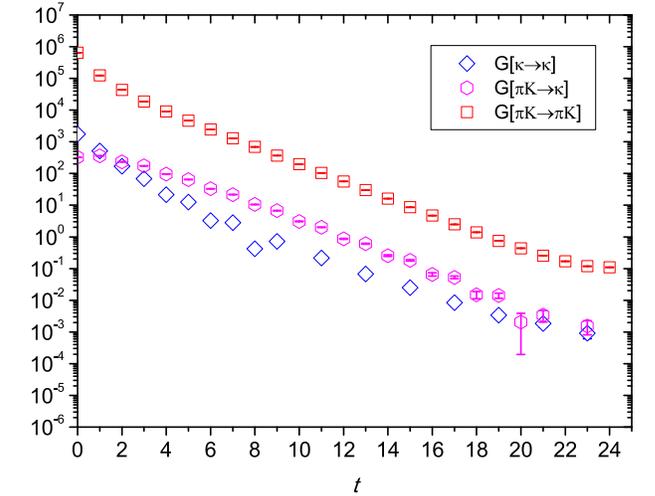}
\end{center}
\vspace{-0.5cm}
\caption{(color online).
Real parts of the diagonal components
($\pi K\to\pi K$ and $\kappa\to\kappa$)
and the real part of the off-diagonal component $\pi K\to\kappa$
of the time correlation function $C(t)$.
Occasional points with negative central values
for the diagonal component $\kappa\to\kappa$ and
the off-diagonal component $\pi K\to\kappa$ are not plotted.
\label{fig:G_t}
}
\end{figure}

We calculate two eigenvalues $\lambda_n(t,t_R)$ ($n=1,2$)
for the matrix $M(t,t_R)$ in Eq.~(\ref{eq:M_def})
with the reference time $t_R = 7$.
In this work, we are only interested in
the eigenvalue $\lambda_1(t,t_R)$~\footnote{
In our previous study~\cite{Fu:2011xw},
we have preliminarily examined the behavior of $\lambda_2(t,t_R)$.
}.
In Figure~\ref{fig:eng_ground}
we plot our lattice results for $\lambda_1(t, t_R)$
for each valence quark $m_x$
in a logarithmic scale as the functions of $t$
together with a correlated fit to the
asymptotic form given in Eq.~(\ref{Eq:asy}).
From these fits we then extract the energies that will
be used to determine the $s$-wave scattering lengths.
\begin{figure*}[th]
\begin{center}
\includegraphics[width=7.5cm]{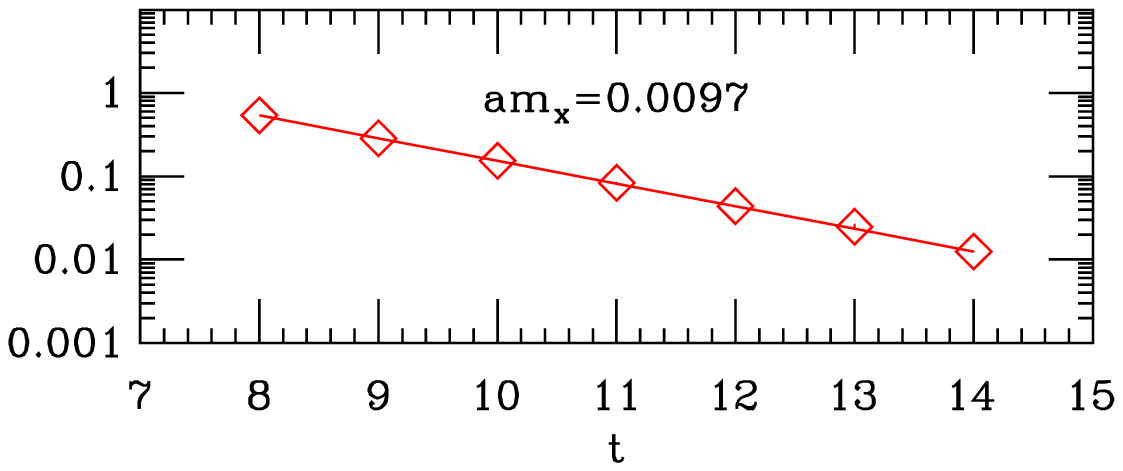}  \hspace{0.8cm}
\includegraphics[width=7.5cm]{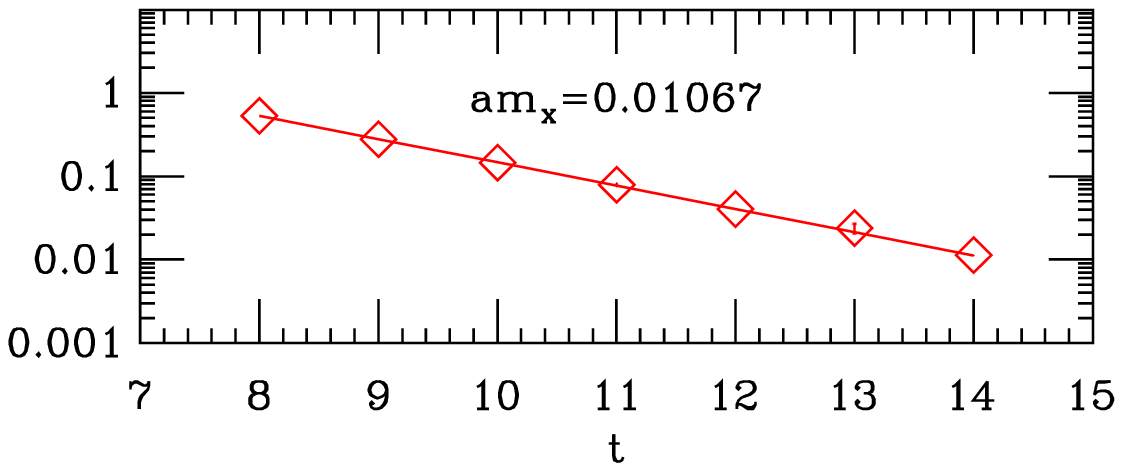}
\includegraphics[width=7.5cm]{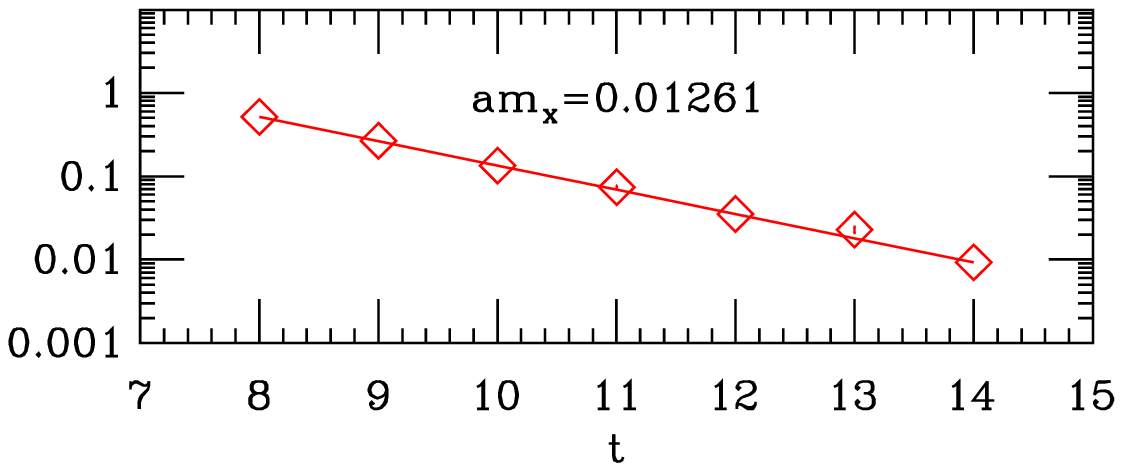}  \hspace{0.8cm}
\includegraphics[width=7.5cm]{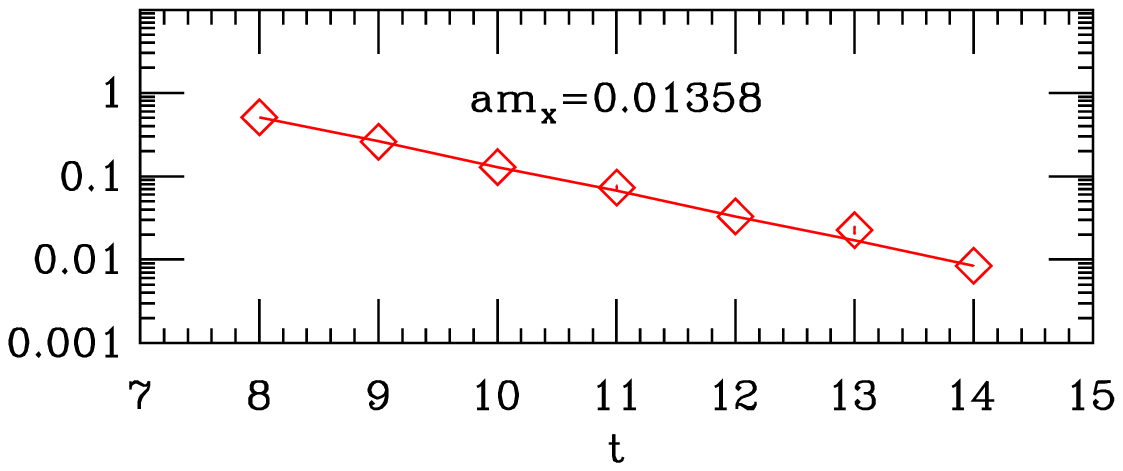}
\includegraphics[width=7.5cm]{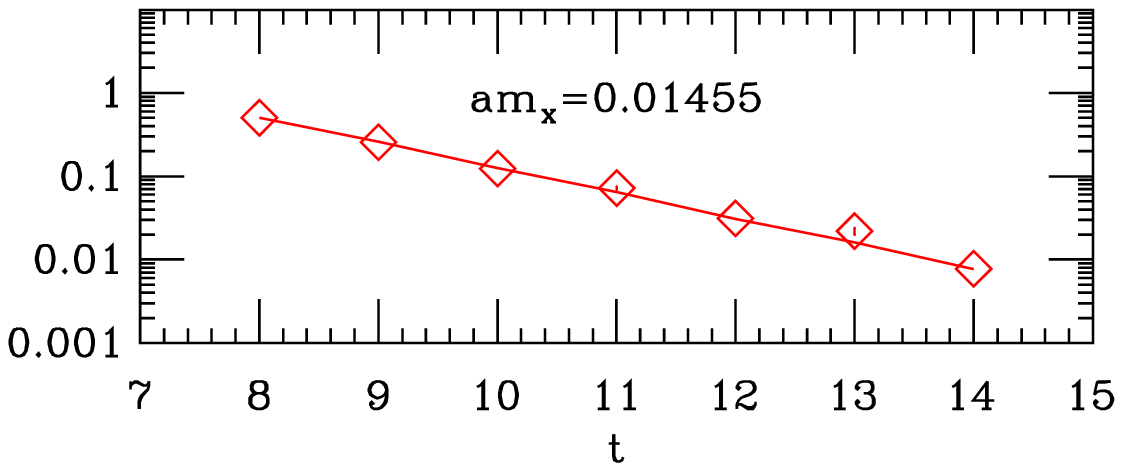}  \hspace{0.8cm}
\includegraphics[width=7.5cm]{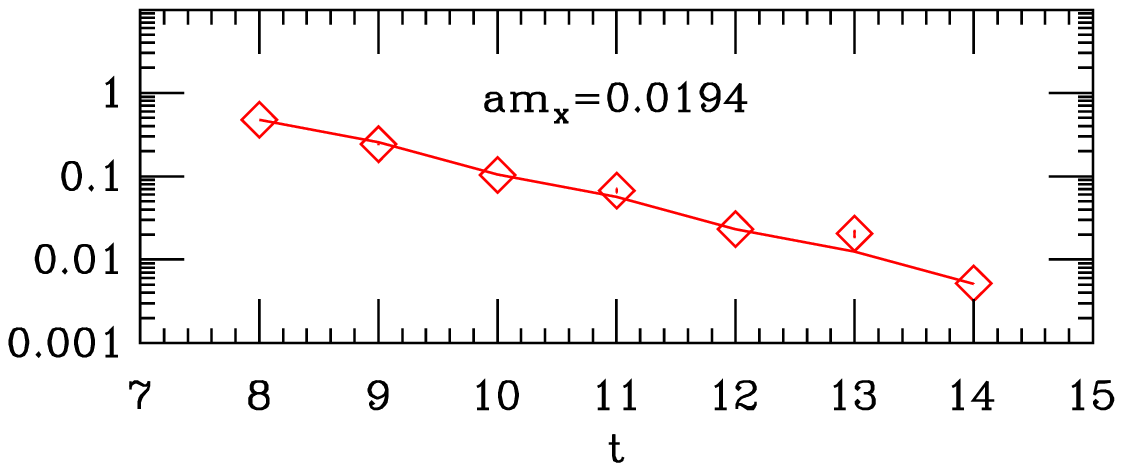}
\end{center}
\vspace{-0.5cm}
\caption{(color online).
The lattice results for $\lambda_1(t, t_R)$ for each valence quark $m_x$
in a logarithmic scale as the functions of $t$
at $I=1/2$ channel are shown.
The solid lines are correlated fits to Eq.~(\ref{Eq:asy}),
from which the energy eigenvalues are extracted.
\label{fig:eng_ground}
}
\end{figure*}

To extract the energies reliably, we must take two major sources
of the systematic errors into consideration.
One arises from the excited states
which affect the correlator in low time slice region.
The other one stems from the thermal contributions
which distort the correlator in high time slice region.
By denoting a fitting range $[t_{\mathrm{min}}, t_{\mathrm{max}}]$
and varying the values of the $t_{\mathrm{min}}$ and $t_{\mathrm{max}}$,
we can control these systematic errors.
In our concrete fitting, we take $t_\mathrm{min}$ to be $t_R+1$ and
increase the reference time slice $t_R$ to
suppress the excited state contaminations.
Moreover, we select $t_\mathrm{max}$ to be sufficiently
far away from the time slice $t=T/2$ to avoid the unwanted thermal contributions.
The fitting parameters $t_R$, $t_{\mathrm{min}}$ and $t_{\mathrm{max}}$
are tabulated in Table~\ref{tab:fitting_results1_2}.
The corresponding fitting results in the reasonable values of $\chi^2/\mathrm{dof}$.
The $\chi^2/\mathrm{dof}$ together with the fit results for
the energies for the ground state  $a E$ are also listed
in Table~{\ref{tab:fitting_results1_2}}.

\begin{table}[h]	
\caption{\label{tab:fitting_results1_2}
Summary of the lattice simulation results for the fitted values of
the energy eigenvalues for the ground state for $I=1/2$ channel.
The third block shows the energy for the ground state in lattice units.
In the table we list the reference time $t_R$, the
lower and upper bound of the fitting range, $t_{\mathrm{min}}$ and
$t_{\mathrm{max}}$, the number of degrees of freedom (dof) for the fit quality $\chi^2/\mathrm{dof}$.
All errors are calculated from jackknife.
}
\begin{ruledtabular}
\begin{tabular}{ccccccl}
Isospin &$a m_x$   & $a E$ & $t_R$ &  $t_{\mathrm{min}}$
& $t_{\mathrm{max}}$  & $\chi^2/{\rm dof}$ \\
\hline
\multirow{6}*{$I=\frac{1}{2}$}
&$0.00970$  & $0.6260(6)$   & $7$  &  $8$  & $14$  & $2.37/3$  \\
&$0.01067$  & $0.6412(7)$   & $7$  &  $8$  & $14$  & $2.83/3$  \\
&$0.01261$  & $0.6694(10)$  & $7$  &  $8$  & $14$  & $3.63/3$  \\
&$0.01358$  & $0.6828(10)$  & $7$  &  $8$  & $14$  & $4.55/3$  \\
&$0.01455$  & $0.6952(11)$  & $7$  &  $8$  & $14$  & $5.57/3$  \\
&$0.01940$  & $0.7561(12)$  & $7$  &  $8$  & $14$  & $13.1/3$  \\
\end{tabular}
\end{ruledtabular}
\end{table}

Now we can insert these energies in Table~\ref{tab:fitting_results1_2}
into the Eqs.~(\ref{eq:exact}) and (\ref{eq:luscher})
to obtain the scattering lengths.
The center-of-mass scattering momentum $k^2$ in GeV
calculated by Eq.~(\ref{eq:MF_k_e})
and thence the corresponding $s$-wave scattering lengths
in lattice units obtained through Eqs.~(\ref{eq:exact}) and (\ref{eq:luscher})
are summarized in Table~\ref{tab:I1_2}.
Here we use the pion masses and kaon masses
given in Table~\ref{tab:m_pi_K}.
The errors of the center-of-mass scattering momentum $k$ and
the $s$-wave scattering lengths are calculated from
the statistic errors of the energy shifts energies $\delta E$,
pion mass $m_\pi$ and kaon mass $m_K$.

\begin{table}[h]	
\caption{\label{tab:I1_2}
Summary of the lattice simulation results of
the $s$-wave scattering lengths for $I=1/2$ channel.
The third block shows the center-of-mass scattering momentum $k^2$ in GeV,
Column four shows the scattering lengths in lattice units,
and Column five shows the pion mass times scattering lengths.
}
\begin{ruledtabular}
\begin{tabular}{ccccl}
Isospin &$am_x$   &  $k^2$[${\rm GeV}^2$]  & $a_0$  & $m_\pi a_0$\\
\hline
\multirow{6}*{$I=\frac{1}{2}$}
&$0.00970$  &  $-0.00884(27)$  &  $2.32(29)$  & $0.543(34)$  \\
&$0.01067$  &  $-0.00904(29)$  &  $2.37(28)$  & $0.588(38)$  \\
&$0.01261$  &  $-0.00956(42)$  &  $2.34(29)$  & $0.690(65)$  \\
&$0.01358$  &  $-0.01010(47)$  &  $2.69(30)$  & $0.778(78)$  \\
&$0.01455$  &  $-0.01075(52)$  &  $2.56(37)$  & $0.887(94)$  \\
&$0.01940$  &  $-0.01343(68)$  &  $2.94(55)$  & $1.498(181)$ \\
\end{tabular}
\end{ruledtabular}
\end{table}

\begin{figure}[thb!]
\begin{center}
\includegraphics[width=8.0cm]{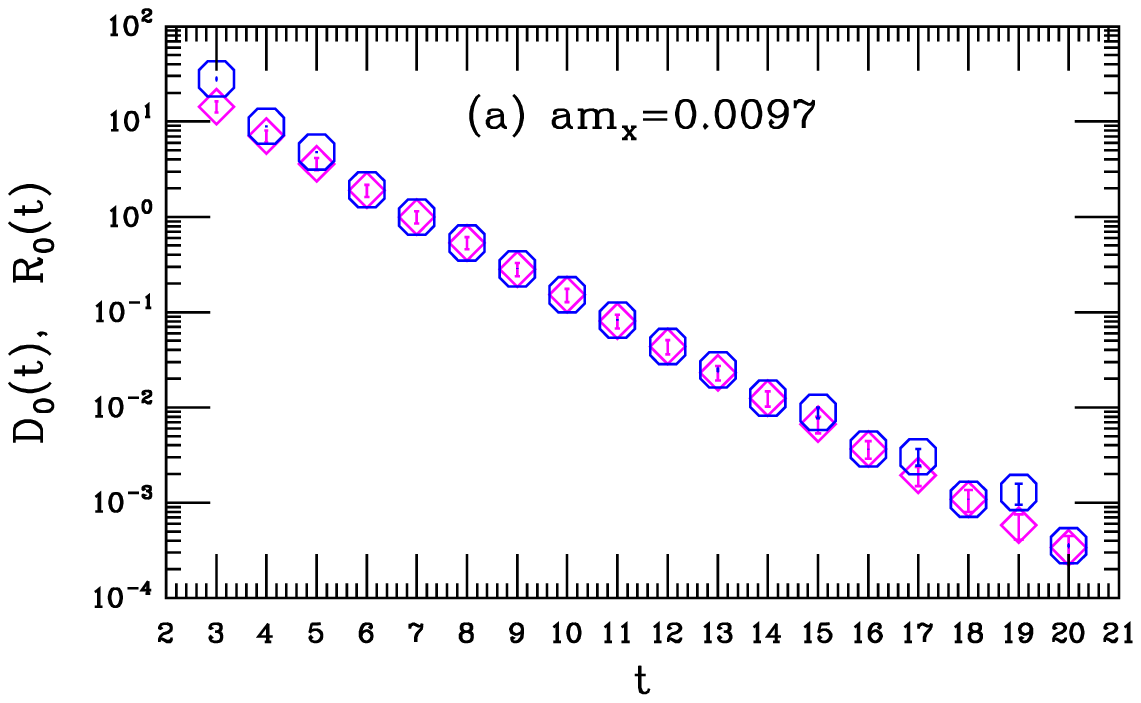}
\includegraphics[width=8.0cm]{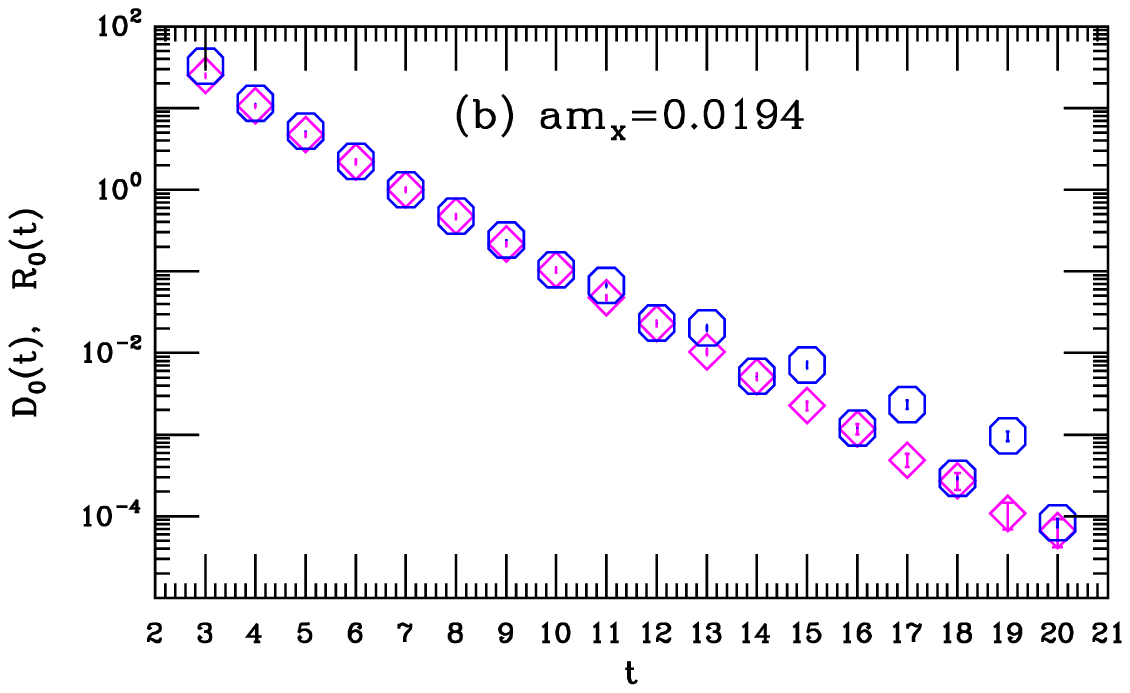}
\end{center}
\vspace{-0.5cm}
\caption{(color online).
$R_0(t)$ (magenta diamonds) and $D_0(t)$ (blue octagons) at
(a) $am_x=0.0097$ or $m_\pi\simeq 0.33$ GeV and
(b) $am_x=0.0194$ or $m_\pi\simeq 0.466$ GeV
for $I=1/2$ channel.
\label{fig:corr-ratio}
}
\end{figure}
To examine the effects of the contaminations from the excited states
for $I=1/2$ channel,
we denote the ratios of $C_{00}(t)$ and
$\mbox{EV}[C^{-1}(t_R) \, C(t) ]_{00}$ which is
the lowest eigenvalue of $C^{-1}(t_R) \, C(t)$,
to the $\pi K$ correlator~\cite{Sasaki:2010zz},
\begin{eqnarray}
R_0(t) &\equiv& \frac{C_{00}(t)}{C_{00}(t_R)}, \cr
D_0(t) &\equiv& \mbox{EV} \left[ C^{-1}(t_R) \, C(t) \right]_{00} \,.
\end{eqnarray}
In the upper panel of Figure~\ref{fig:corr-ratio},
we show $R_0(t)$ (magenta diamonds) and $D_0(t)$ (blue octagons)
at $m_\pi\simeq 0.33$ GeV.
We can note that the difference of two ratios is small.
This suggests that the contamination from the excited states
is negligible at this light quark mass.
However, from the bottom panel of Figure~\ref{fig:corr-ratio},
we observe that the contamination from the excited states
is not negligible at $m_\pi\simeq 0.466$ GeV,
and the diagonalization obviously
changes the characteristics of the ratio,
since the $\pi K$ interpolative operator for $I=1/2$ channel
has a large overlap with the excited states~\cite{Sasaki:2010zz}.
Therefore, we further confirmed that
the separation of the contamination from the excited states is
absolutely necessary for the heavy quark masses~\cite{Sasaki:2010zz}
when we study the $\pi K$ scattering for $I=1/2$ channel.

Using the fitting models discussed in Ref.~\cite{Aubin:2004wf},
we extract the pion and kaon masses~\cite{fzw:2011cpl}.
And using the fitting model in Eq~(\ref{eq:kfit}),
we calculate the kappa mass~\cite{Fu:2011xb}.
In Figure~\ref{fig:kappa_limit},
we display $m_\kappa, m_K, m_\pi$, and $\pi K$ threshold $m_{\pi+K}$
in lattice units as the functions of the pion mass $m_\pi$.
We observe that, as the valence quark mass increases,
$\pi K$ threshold grows faster than $\kappa$ mass and,
as a consequence, for the heavy quark masses,
$\pi K$ threshold is close to the $\kappa$ mass.
In Figure~\ref{fig:kappa_limit}, we can clearly note that
$\pi K$ threshold is very close to the $\kappa$ mass
for light quark mass $a m_x = 0.0097$.
This can be in part explained why
the separation of the contamination from the excited states is
indispensable for the large quark masses~\cite{Sasaki:2010zz}.
\begin{figure}[thb!]
\begin{center}
\includegraphics[width=8.0cm]{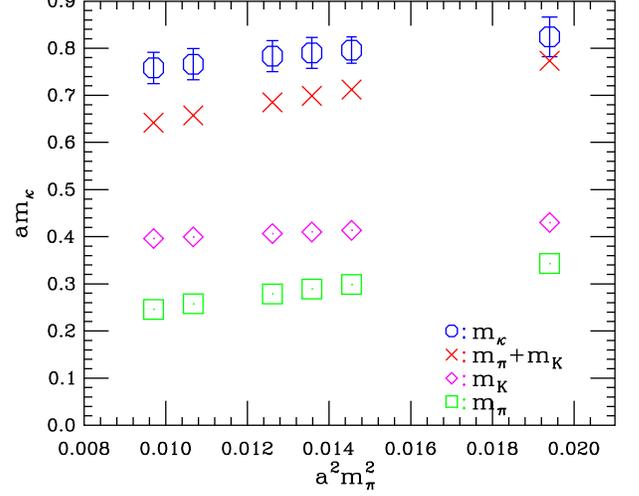}
\end{center}
\caption{(color online). \label{fig:kappa_limit}
Characteristics of $m_\kappa, m_K, m_\pi$ and $m_{\pi+K}$
in lattice units as the functions of the pion mass.
}
\end{figure}

\subsection{Chiral extrapolations and scattering length}
In the present study, we employ a reasonable small pion masses $m_\pi$,
namely, $m_\pi = 0.330 - 0.466$ GeV,
which are still considerably larger than its physical ones,
therefore, we need to extrapolate the $s$-wave $\pi K$ scattering lengths
toward the physical point.
For this purpose, we employ the formula predicted by
$\rm SU(3)$ chiral perturbation theory to next-to-leading order (NLO)~\cite{Bernard:1990kw,Kubis:2001bx,Beane:2006gj,Sasaki:2010zz,Chen:2006wf}.
In $\rm SU(3)$ chiral perturbation theory at NLO,
we provide the continuum $\rm SU(3)$ $\chi$PT forms of
$a_0^{I=\frac{3}{2}}$ and $a_0^{I=\frac{1}{2}}$,
which can be directly constructed
from Refs.~\cite{Kubis:2001bx,Beane:2006gj,Sasaki:2010zz},
\begin{widetext}
\begin{eqnarray}
\label{eq:a0_alg}
a_0^{I=\frac{3}{2}} &=& \frac{\mu_{\pi K}}{4\pi f_\pi^2}
\left\{ - 1 +  \frac{ 32m_\pi m_K}{f_\pi^2} L_{\pi K}(\Lambda_\chi)-
\frac{ 16 m_\pi^2}{f_\pi^2} L_{5}(\Lambda_\chi)+
\frac{1}{16\pi^2 f_\pi^2} \chi_{\pi K}^{I=\frac{3}{2}}(\Lambda_\chi,m_\pi,m_K)
\right\} , \cr
a_0^{I=\frac{1}{2}} &=& \frac{\mu_{\pi K}}{4\pi f_\pi^2}
\left\{ 2 + \frac{ 32 m_\pi m_K}{f_\pi^2} L_{\pi K}(\Lambda_\chi) +
\frac{ 32 m_\pi^2}{f_\pi^2} L_{5}(\Lambda_\chi)+
\frac{1}{16\pi^2 f_\pi^2} \chi_{\pi K}^{I=\frac{1}{2}}(\Lambda_\chi,m_\pi,m_K)
\right\} ,
\end{eqnarray}
\end{widetext}
where we plugged in the pion mass $m_\pi$, the kaon mass $m_K$,
and the pion decay constant $f_\pi$,
which are summarized Table~\ref{tab:m_pi_K}.
$L_5(\Lambda_\chi)$ and
$L_{\pi K}(\Lambda_\chi) \equiv 2L_1 + 2L_2 + L_3 - 2L_4 - L_5/2 + 2L_6 + L_8$ are
low energy constants defined in Ref.~\cite{Gasser:1984gg}
at the chiral symmetry breaking scale $\Lambda_\chi$.
We should bear in mind that the expressions in Eq.~(\ref{eq:a0_alg})
are written in terms of full $f_\pi$, and not its chiral limit value.
The $\chi_{\pi K}^{I=\frac{3}{2}}(\Lambda_\chi,m_\pi,m_K)$ and
$\chi_{\pi K}^{I=\frac{1}{2} }(\Lambda_\chi,m_\pi,m_K)$
are the known functions at NLO which clearly
depend upon the chiral scale $\Lambda_\chi$
with chiral logarithm terms, namely,
\begin{widetext}
\begin{eqnarray}
\chi_{\pi K}^{I=\frac{3}{2}}(\Lambda_\chi,m_\pi,m_K) &=&
\kappa_\pi  \ln \frac{m_\pi^2 }{\Lambda_\chi^2} +
\kappa_K    \ln \frac{m_K^2   }{\Lambda_\chi^2} +
\kappa_\eta \ln \frac{m_\eta^2}{\Lambda_\chi^2} +
\frac{86}{9} m_K m_\pi +
\kappa_{\tan} \arctan \left( \frac{2(m_K -m_\pi)}{m_K +2m_\pi}
\sqrt{ \frac{m_K + m_\pi }{ 2m_K -m_\pi }} \right) , \cr
\chi_{\pi K}^{I=\frac{1}{2}}(\Lambda_\chi,m_\pi,m_K) &=&
\kappa_{\pi }^{\prime} \ln \frac{m_\pi^2 }{\Lambda_\chi^2} +
\kappa_{K   }^{\prime} \ln \frac{m_K^2   }{\Lambda_\chi^2} +
\kappa_{\eta}^{\prime} \ln \frac{m_\eta^2}{\Lambda_\chi^2} +
\frac{86}{9} m_K m_\pi +
\frac{3}{2}\kappa_{\tan}\arctan
\left( \frac{2(m_K -m_\pi)}{m_K +2m_\pi}
\sqrt{ \frac{m_K + m_\pi }{ 2m_K -m_\pi }} \right) \cr
&+&
\kappa_{\tan}^{\prime} \arctan
\left( \frac{2(m_K + m_\pi)}{m_K - 2m_\pi}
\sqrt{ \frac{m_K - m_\pi }{ 2m_K + m_\pi }} \right) ,
\end{eqnarray}
\end{widetext}
with
\begin{eqnarray}
\kappa_\pi &=&
\frac{11m_Km_\pi^3 + 8m_\pi^2m_K^2 -5m_\pi^4}{2(m_K^2 -m_\pi^2)} , \cr
\kappa_K   &=&
-\frac{67m_K^3 m_\pi - 8m_\pi^3m_K + 23m_K^2m_\pi^2}{9(m_K^2-m_\pi^2)} , \cr
\kappa_\eta &=&
\frac{24m_\pi m_K^3 -5 m_K m_\pi^3 + 28m_K^2 m_\pi^2 - 9m_\pi^4}
     {18(m_K^2-m_\pi^2)} , \cr
\kappa_{\rm tan} &=&
-\frac{16 m_Km_\pi}{9}\frac{\sqrt{2m_K^2 +m_Km_\pi -m_\pi^2}}{m_K-m_\pi} , \cr
\kappa_{\pi}^{\prime}  &=&
\frac{11m_Km_\pi^3-16m_K^2m_\pi^2+10m_\pi^4}{2(m_K^2-m_\pi^2)} , \cr
\kappa_{K}^{\prime}    &=&  -
\frac{67 m_{K}^3 m_\pi-8 m_{\pi}^3 m_K-46 m_{K}^2 m_{\pi}^2}{9(m_K^2-m_\pi^2)} , \cr
\kappa_{\eta}^{\prime} &=&
\frac{24m_\pi m_K^3-5m_Km_\pi^3-56m_K^2m_\pi^2+18m_\pi^4}{18(m_K^2-m_\pi^2)} , \cr
\kappa_{tan}^{\prime}  &=&
\frac{8 m_Km_\pi}{9}\frac{\sqrt{2m_K^2 -m_Km_\pi -m_\pi^2}}{m_K+m_\pi} .
\end{eqnarray}
%
\begin{figure}[h]
\includegraphics[width=80mm]{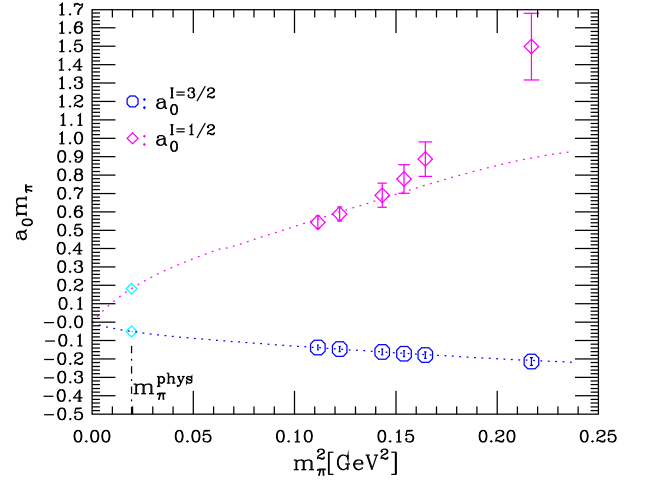}
\caption{\label{fig:ChPT-fit}
$m_{\pi}^2$-dependence of the $\pi K$ scattering lengths $m_{\pi} a_0$
for both $I=1/2$ and $I=3/2$ channels.
The dotted lines give the $\rm SU(3)$ $\chi$PT predictions at NLO.
The cyan diamond point indicate its physical values.
}
\end{figure}

In this work, we did not measure the $\eta$ mass ($m_\eta$),
alteratively,  we utilize the Gell-Mann-Okubo mass-relation
to determine $\eta$-mass.
To improve the $\chi$PT fit, in principle,
we can include all the lattice simulation data of
the $\pi K$ scattering lengths for both $I=1/2$ and $3/2$ channels
to perform the simultaneous fitting.
However, the fit with the data of the scattering lengths
for $I=1/2$ channel in $m_\pi \ge 0.392$ GeV
significantly increases $\chi^2/\mathrm{dof}$,
so we only use the scattering lengths  for $I=1/2$ channel
in $m_\pi < 0.392$ GeV.
The fitting results of $\pi K$ scattering lengths,
$m_\pi a_0^{I=3/2} $ and $m_\pi a_0^{I=1/2}$
are plotted by the dotted lines as the functions of $m_\pi^2$
in Figure~\ref{fig:ChPT-fit}.
The dotted lines are the chiral extrapolation of
the scattering lengths for both isospin eigenstates.
The fit parameters $L_{\pi K}, L_5$, and the scattering lengths
$m_\pi a_0$ at the physical points (namely, $m_\pi=0.140$ GeV, $m_K=0.494$ GeV)~\cite{Nakamura:2010zzi}
are also summarized in Table~\ref{tb:ChPT-fit},
where the chiral scale $\Lambda_\chi$ is taken as
the physical $\eta$ mass, namely, $\Lambda_\chi = 0.548$ GeV~\cite{Nakamura:2010zzi}
as it is done in Ref.~\cite{Aubin:2004fs}.
The cyan diamond points in Figure~\ref{fig:ChPT-fit}
shows the values of the physical scattering lengths.
From Figure~\ref{fig:ChPT-fit}, we note that
our lattice simulation results for $I = 3/2$ scattering length
agrees well with the one-loop formula,
while scattering lengths for $I = 1/2$ channel have a large error,
and is in reasonable agreement
with the $\rm SU(3)$ $\chi$PT at NLO.

\begin{table}[h]
\caption{\label{tb:ChPT-fit}
The fitted $s$-wave scattering lengths $m_\pi a_0$
at the physical point ($m_\pi=0.140$ GeV, $m_K=0.494$ GeV).
The chiral scale $\Lambda_\chi$ is taken as the physical $\eta$ mass.
}
\begin{ruledtabular}
\begin{tabular}{cccccc}
$\chi^2/{\mathrm{dof}}$  & $10^3 \cdot L_{\pi K}$  & $10^3 \cdot L_5$   &
$m_\pi a_0^{ I=\frac{3}{2} }$  & $m_\pi a_0^{ I=\frac{1}{2} }$    \\
\hline
$ 0.287/7 $  & $ 1.48(5)$  &   $1.06(14)$  & $-0.0505(19)$  & $ 0.1827(37)$  \\
\end{tabular}
\end{ruledtabular}
\end{table}

The fitted value of the $L_5$ is reasonable consistent with the value
evaluated by PACS-CS collaboration~\cite{Aoki:2008sm},
and is smaller than the corresponding result
evaluated by MILC collaboration~\cite{Aubin:2004fs}
and NPLQCD collaboration~\cite{Beane:2006gj}.
The fitted value of $L_{\pi K}$ is also smaller than
the result evaluated by NPLQCD collaboration~\cite{Beane:2006gj}.
The $s$-wave $\pi K$ scattering lengths for
both $I=1/2$ and $3/2$ channels are in agreement with
the other lattice studies~\cite{Miao:2004gy, Beane:2006gj,
Nagata:2008wk, Sasaki:2010zz}.

\section{Summary and outlook}
\label{sec:conclude}
In the present study, we carried out a direct lattice QCD calculation
of the $s$-wave $\pi K$ scattering lengths
for both isospin $I = 1/2$ and $3/2$ channels,
where the rectangular diagram plays a crucial role,
for the MILC ``medium'' coarse ($a=0.15$ fm) lattice
ensemble in the presence of $2+1$ flavors of the Asqtad
improved staggered dynamical sea quarks,
generated by the MILC collaboration.
We employed almost same technique in Ref.~\cite{Kuramashi:1993ka}
but with the moving wall sources without gauge fixing
to obtain the reliable precision.
We calculated all the three diagrams
categorized in Ref.~\cite{Nagata:2008wk},
and observed a clear signal of the attraction for $I=1/2$ channel and
that of repulsion for $I=3/2$ channel, respectively.
Moreover, for $I = 1/2$ channel,
we employed the variational method
to isolate the contamination from the excited states.
We further confirmed that the separation of the contamination is
absolutely necessary for the heavy quark masses~\cite{Sasaki:2010zz}
when we study $\pi K$ scattering in $I=1/2$ channel.
Simultaneously extrapolating our lattice simulation data of
the $s$-wave scattering lengths for both isospin eigenstates
to the physical point gives the scattering lengths
$m_\pi  a_{3/2} = -0.0505(19)$ and  $m_\pi  a_{1/2} = 0.1827(37)$
for the $I=3/2$ and $1/2$ channels, respectively,
which are in accordance with the current theoretical predictions
to one-loop levels and the present experimental reports,
and can be comparable with the other lattice
studies~\cite{Miao:2004gy, Beane:2006gj, Nagata:2008wk, Sasaki:2010zz}.

A clear signal can be seen for long time separation range
in the rectangular diagram of the $\pi K$ scattering.
Reducing the noise by performing the calculation on a larger volume
or the smaller pion mass could further improve the signal to
noise ratio for the rectangular diagram,
and therefore obtain better results
for the scattering length in $I = 1/2$ channel~\cite{Sasaki:2010zz}.
Moreover, the behavior near the chiral limit is strongly affected
by the chiral logarithm term, so giving an evaluation
without the long chiral extrapolation is highly desirable
to ensure the convergence of the chiral expansion~\cite{Sasaki:2010zz}.
Furthermore, $\tan\delta_0(k)/k$ in the low-momentum limit
must be evaluated by the systematic studies with
the different volumes and boundary conditions~\cite{Sasaki:2010zz}.
For these purposes, we are planning a series of lattice simulations
on MILC coarse, fine, and superfine lattice ensembles
with concentrating on the lightest accessible values of
the quark masses, namely, in $m_\pi< 300$ MeV.
We anticipate that these future tasks
should make the calculation of the rectangular diagram more reliable.

It is well-known that $\pi K$ scattering at $I=1/2$ channel is
a more challenging and interesting channel phenomenologically
due to the existence of kappa resonance.
The study of the $s$-wave $\pi K$ scattering at zero momentum is
just first step in the study of
the hadron interactions including $s$-quarks.
However, it is particularly encouraging that $\pi K$ scattering
for $I=1/2$ channel can be reliably calculated
by the moving wall sources without gauge fixing in spite of
the essential difficulties of the calculation of
the four-point functions, especially rectangular diagram.
It raises a prospect that this technique can be successfully
employed to investigate the $\kappa$ resonance, and so on.

In our previous work~\cite{fzw:2011cpc12},
we have precisely evaluated the $\kappa$ mass,
and found that the decay $\kappa \to \pi K$ is
only allowed kinematically for enough small $u$ quark mass.
This work and our preliminary lattice simulation reported here
for $\pi K$ scattering lengths
will encourage the researchers to study the $\kappa$ resonance.
We are beginning a series of lattice investigations on
the $\kappa$ resonance parameters
with isospin representation of $(I, I_z) = (1/2, 1/2)$,
and the preliminary results are already reported
in Refs.~\cite{Fu:2011xw,Fu:2011xz}.

\section*{Acknowledgments}
This work is supported in part by Fundamental Research Funds
for the Central Universities (2010SCU23002) and
the Startup Grant from the Institute of Nuclear Science
and Technology of Sichuan University.
The author thanks Carleton DeTar for kindly providing us
the MILC gauge configurations used for this work and
the fitting software to analyze the lattice data.
We are indebted to the MILC collaboration
for using the Asqtad lattice ensemble and MILC codes.
We are grateful to Hou Qing for his comprehensive supports.
The computations for this work were carried out at AMAX,
CENTOS and HP workstations in the Institute of
Nuclear Science and Technology, Sichuan University.

\appendix
\section{The calculation method of zeta function }
\label{appe:zeta}
In this appendix we briefly discuss one method for
numerical evaluation of zeta function $\mathcal{Z}_{00}(s;q^2)$
in the center-of-mass system for any value of $q^2$.
Here we follow the original derivations and notations in Ref.~\cite{Yamazaki:2004qb}.

The definition of zeta function $\mathcal{Z}_{00}(s;q^2)$
in Eq.~(\ref{eq:Z00d}) is
\begin{equation}
\sqrt{4\pi} \cdot \mathcal{Z}_{00}(s;q^2) =
\sum_{{\mathbf n}\in\mathbb{Z}^3}\frac{1}{(n^2-q^2)^{s}}\,.
\end{equation}
The zeta function $\mathcal{Z}_{00}(s; q^2)$ takes a finite value
for ${\mathrm Re}\, s > 3/2 $,
and $\mathcal{Z}_{00}(1; q^2)$,
is defined by the analytic continuation from the region ${\mathrm Re}\,s > 3/2$.

First we consider the case of $q^2 > 0$,
and we separate the summation in $\mathcal{Z}_{00}( s ; q^2 )$ into two parts as
\begin{equation}
\sum_{{\mathbf n}\in\mathbb{Z}^3 } \frac{1}{( n^2 - q^2 )^{s}} =
\sum_{n^2 < q^2}\frac{1}{(n^2-q^2)^{s}} +
\sum_{n^2 > q^2}\frac{1}{(n^2-q^2)^{s}} \,,
\label{eq:app_zeta_1}
\end{equation}
The second term can be written in an integral form,
\begin{widetext}
\begin{eqnarray}
\sum_{ n^2 > q^2 } \frac{1}{( n^2 - q^2 )^{s}} & = &
\frac{ 1 }{ \Gamma ( s ) } \sum_{ n^2 > q^2 }
\left[ \int_0^1        {\rm d}t \ t^{s-1} e^{ - t ( n^2 - q^2) } +
       \int_1^{\infty} {\rm d}t \ t^{s-1} e^{ - t ( n^2 - q^2) } \right] \cr
& = &
\frac{ 1 }{ \Gamma(s) }
\int_0^1 {\rm d}t \ t^{s-1} e^{q^2 t} \sum_{{\mathbf n}\in\mathbb{Z}^3} e^{-n^2 t} -
\sum_{ n^2 < q^2 } \frac{1}{( r^2 - q^2 )^{s}} +
\sum_{{\mathbf n}\in\mathbb{Z}^3} \frac{ e^{ -( n^2 - q^2 ) } }{ ( n^2 - q^2 )^s }\,.
\label{eq:app_zeta_2}
\end{eqnarray}
\end{widetext}
The second term neatly cancels out the first term in Eq.~(~\ref{eq:app_zeta_1}).
Next we rewrite the first term in Eq.~(\ref{eq:app_zeta_2})
by the Poisson's resummation formula as,
\begin{eqnarray}
\hspace{-1.0cm} \frac{1}{\Gamma(s)} \int_0^1 \ {\rm d}t \ t^{s-1} e^{ t q^2 }
\sum_{{\bf n } \in \mathbb{Z}^3} e^{-n^2 t} &=&  \cr
\hspace{-4.0cm} && \hspace{-4.0cm} \frac{1}{ \Gamma (s) }
\int_0^1  {\rm d}t \ t^{ s - 1 } e^{ t q^2 }
\left( \frac{\pi}{t} \right)^{3/2}
\sum_{{\bf n}\in \mathbb{Z}^3} e^{-\pi^2 n^2/t} \,.
\label{eq:app_zeta_3}
\end{eqnarray}
The divergence at $s=1$ comes from the ${\bf n} = {\bf 0}$ part of the integrand
on the right-hand side, therefore we divide the integrand into
a divergent part (${\bf n} = {\bf 0}$) and a finite part (${\bf n} \ne {\bf 0}$).
The divergent part can be evaluated for $ {\mathrm Re}\,s > 3 / 2 $ as
\begin{equation}
\int_0^1 \ {\rm d}t \ t^{s-1}
e^{q^2 t} \left( \frac{ \pi }{ t } \right)^{ 3 / 2 } =
\sum_{l=0}^{ \infty }\frac{ \pi^{3/2}}{s+l-3/2}\frac{q^{2l}}{ l ! }\,.
\label{eq:app_zeta_4}
\end{equation}
The right hand side of this equation takes a finite value at $s=1$.

After gathering all terms we obtain the representation of
the zeta function in the center-of-mass system at $s=1$,
\begin{widetext}
\begin{eqnarray}
\sqrt{4\pi} \cdot \mathcal{Z}_{00}(s;q^2) & = &
\sum_{ {\bf n } \in \mathbb{Z}^3 }
\frac{e^{-(n^2 - q^2)}}{ n^2 - q^2} + \sum_{ l = 0 }^{ \infty }
\frac{ \pi^{3/2}}{l-1/2} \frac{q^{2l} }{ l ! } + \int_0^1  {\rm d}t \ e^{q^2 t}
\left( \frac{ \pi }{ t } \right)^{3/2}
\sum_{{\bf n} \in \mathbb{Z}^3}\!\!\rule{0mm}{1em}^{\prime} e^{-\pi^2 n^2/t } \,,
\label{eq:app_zeta_s=1}
\end{eqnarray}
\end{widetext}
where $\sum_{ {\bf n} \in \mathbb{Z}^3 }^{\prime}$ stands for
a summation without ${\bf n} = {\bf 0}$.

For the case of $q^2 \le 0$,
it is not necessary for us to separate the summation
in $\mathcal{Z}_{00}( s ; q^2 )$,
and it can be also written in an integral form,
\begin{eqnarray}
\hspace{-1.0cm}  \sum_{ {\mathbf n}\in\mathbb{Z}^3 } \frac{1}{( n^2 - q^2 )^{s}} & = &\nonumber \\
\hspace{-3.0cm} && \hspace{-3.0cm} \frac{ 1 }{ \Gamma(s) }
\int_0^1 {\rm d}t \ t^{s-1} e^{q^2 t} \sum_{{\mathbf n}\in\mathbb{Z}^3} e^{-n^2 t} +
\sum_{{\mathbf n}\in\mathbb{Z}^3} \frac{ e^{ -( n^2 - q^2 ) } }{ ( n^2 - q^2 )^s }\,.
\end{eqnarray}
Following the same procedures, we arrive at the same expression
in Eq.~(\ref{eq:app_zeta_s=1}).
Hence, Eq.~(\ref{eq:app_zeta_s=1}) can be applied for both cases.

We also note that, for negative $q^2$, an exponentially convergent expression
of the zeta function $\mathcal{Z}_{00}( s ; q^2 )$ has been derived in
Ref.~\cite{neg}. We numerically compared
this representation of the zeta functions with that of above described representation,
and found agreement.
Therefore, we use Eq.~(\ref{eq:app_zeta_s=1}) in this work.


\end{document}